\documentclass[useAMS,usenatbib,usegraphicx,referee]{mn2e}
\begin{document}

\title{Criteria for mixing rules application for inhomogeneous astrophysical grains}
\author[N.Maron and O. Maron]{N. Maron\thanks{E-mail:
N.Maron@if.uz.zgora.pl; } and O.Maron$^{1}$\\
$^{1}$J. Kepler Institute of Astronomy, University of Zielona G\'ora,
              ul. Lubuska 2, 65-265 Zielona G\'ora, Poland \\}

\date{Received ; accepted }
\pagerange{\pageref{firstpage}--\pageref{lastpage}} \pubyear{2008}

\maketitle

\label{firstpage}

\begin{abstract}
The analysis presented in this paper verifies which of the mixing rules are best for real components of interstellar dust in possible wide range of wavelengths.The DDA method with elements of different components with various volume fractions has been used. We have considered 6 materials: ice, amorphous carbon, graphite, SiC, silicates and iron, and the following mixing rules: Maxwell-Garnett, Bruggeman, Looyenga, Hanay and Lichtenecker which  must satisfy rigorous bounds. The porous materials have also been considered. We have assumed simplified spatial distribution, shape and size of inclusions. The criteria given by \citet{draine1988} have been used to determine the range of wavelengths for the considered mixtures in order to calculate the ${\rm Q_{ext}}$ using the DDA. From all chosen mixing rules for the examined materials in majority of cases (13 out of 20) the best results have been obtained using the Lichtenecker mixing rule. In 5 cases this rule is better for some volume fraction of inclusions.

\end{abstract}

\begin{keywords}
ISM: general, dust, extinction, interstellar grains, mixing rules, discrete dipole approximation
\end{keywords}

\section{Introduction}
Interstellar dust is a mixture of grains of different shape, size and chemical composition. The most frequent are assumed to be different allotropic forms of carbon, ${\rm \alpha}$-SiC, "astronomical silicate" and ice. Many authors proposed grains which have been mixtures of various materials. In order to obtain the refractive indices of such types of grains the formulae describing different mixing rules are used. Those formulae have been derived for various assumptions concerning arrangement of inclusions and their shape. The best known examples of the effective medium theories (EMT) are theories by Maxwell-Garnett and Bruggeman. The Maxwell-Garnett mixing rule has been re-derived by Bohren and Wickramasinghe \citep{bohren1977} for the spherical inclusions arranged chaotically. The assumption of spherical shape of inclusions or generally separated inclusion structure causes asymmetry of this mixing rule, whereas the formula derived by Bruggeman (and a similar one by Landauer) with non-spherical inclusions, tightly adjoining to each other aggregate structures leads to full symmetry \citep{chylek1983}. Also in deriving the Looyenga and Hanay rules particular models of mixtures were used (\citet{looyenga1965}, \citet{beek1967}, \citet{sihvola1999}). Many existing mixing rules were described by \citet{beek1967} or \citet{sihvola1999}, for example. The Lichtenecker mixing rule has one drawback that it was derived on the basis of fitting to the empirical data. It lacks a physical model apart from some theoretical justification bound with an artificial decomposition of geometrical shapes of inclusions (\citet{zakri}). It is worth mentioning that the interaction between inclusions for small volume fractions is limited and may be omitted. On the other hand for large volume fractions of inclusions this effect is not negligible. Those interactions or their lack are included in more or less explicit way in the assumptions for the mixing rules. Therefore, they lead to the limited applicability of a particular mixing rule. The influence of interactions between inclusions was discussed by \citet{perrin1990} for Maxwell-Garnett and Bruggeman mixing rules. In case of derivation of Looyenga rule the author (\citet{looyenga1965}) avoided the discussion on those interactions. Most of the mixing rules may be applied with good approximation for various components with small volume fraction of inclusions (up to a few per cent) when the average distances between inclusions are large and the interactions are small. For higher volume fractions the choice of a mixing rule is difficult and very important. We used the DDA method with elements (dipoles) of different components (refractive indices) with various volume fractions in order to calculate the extinction coefficients for grains. Many authors (\citet{chylek2000}, \citet{iati2004}, \citet{voshchin2007}) compared the extinction coefficients calculated in this way with the extinction coefficients calculated from Mie theory for grains of refractive indices computed using various mixing rules in order to choose the best rule. The extinction coefficients obtained from the Mie theory are the same as those calculated with the DDA method using a large number of dipoles. Using a large number of dipoles decreases the influence of granularity of grains but at the same time it requires a very long computing time. In order to avoid the influence of the method for calculating the extinction and shorten the computing time we used the DDA method for both cases. Many authors (cf. \citet{beek1967} and \citet{sihvola1999}) compared experimentally obtained refractive indices of mixtures with results given by various mixing rules.Depending on the volume fractions of inclusions, their kinds, shape, spatial distribution and frequency range different mixing rules fitted experimental permittivity. Our choice of mixing rules is based on numerical experiment which allowed us to examine the mixtures in a wide frequency range. In this paper we have assumed random spatial distribution, pseudospherical shape and one size of inclusions. However, in this numerical experiment using the DDA method and the Rayleigh and non-Rayleigh cluster dipol inclusions method (\citet{wolff1994}) we could change the geometrical parameters of inclusions and their spatial distribution according to assumptions leading to different mixing rules and verify their applicability.

\section{Preparation of optical data}

In this work 6 materials have been considered: ice, amorphous carbon, graphite, SiC, silicates and iron. The refractive indices of ice have been taken from \citet{warren1984}, of amorphous carbon from \citet{zubko1996}, for graphite, SiC and "astronomical silicate" from Draine (http://www.astro.princeton.edu/$\sim$draine/dust/dust.diel.html ). The refractive indices of iron have been compiled by \citet{lynch1991}. We have verified whether the Kramers-Kr\"{o}nig relation is fulfilled. The differences of values for n taken from the cited literature, except for iron, and those calculated from Kramers-Kr\"{o}nig relation are negligibly small. For iron there are significant differences due to different methods used by various authors for different wavelength ranges. In order to obtain a homogeneous data sets of refractive indices the values of n have been taken from \citet{lynch1991} and used to calculate k values from Kramers-Kr\"{o}nig relation.
We have interpolated 100 values of n and k in the range from ${\rm 0.0443 \div 150 \mu m}$ and therefore obtained the same wavelength range for further calculations.

\section{Mixing rules for two constituents}
Our study has been limited to grains without electric charge, magnetic susceptibility and only two component mixtures. We have studied 5 mixing rules described in detail in \citet{maron2005} excluding the Rayleigh mixing rule modified by \citet{meredith1960} because this rule has been derived for ordered mixtures \citep{sihvola1999}. The following rules have been taken into account:\\
Asymmetrical
\begin{itemize}
\renewcommand{\labelitemi}{$ $}
\item Maxwell-Garnett \citep{bohren1983}
\begin{equation}
\varepsilon =\varepsilon _{m}+3f\varepsilon _{m}\frac{\varepsilon
_{i}-\varepsilon _{m}}{\varepsilon _{i}+2\varepsilon _{m}-f\left(
\varepsilon _{i}-\varepsilon _{m}\right) },
\end{equation}

\item Hanai-Bruggeman (called Hanai in this paper) \citep{beek1967}
\begin{equation}
\frac{\varepsilon _{i}-\varepsilon }{\varepsilon _{i}-\varepsilon _{m}}
\left( \frac{\varepsilon _{m}}{\varepsilon }\right) ^{\frac{1}{3}}=1-f,
\end{equation}

Symmetrical
\item Bruggeman \citep{bohren1983}
\begin{equation}
f\frac{\varepsilon _{i}-\varepsilon }{\varepsilon
_{i}+2\varepsilon }+(1-f)\frac{\varepsilon
_{m}-\varepsilon }{\varepsilon _{m}+2\varepsilon }=0,
\end{equation}

\item Looyenga \citep{looyenga1965}
\begin{equation}
\varepsilon ^{\frac{1}{3}}=f\mathit{\,}\varepsilon _{i}^{\frac{1}{3}%
}+\left( 1-f\right) \varepsilon _{m}^{\frac{1}{3}},
\end{equation}

\item Lichtenecker \citep{lichtenecker1926}
\begin{equation}
\log \varepsilon =f\log \varepsilon _{i}+\left( 1-f\right)\log \varepsilon _{m}.
\end{equation}

\end{itemize}

In all formulae $f$ is the volume fraction of inclusions and $\varepsilon _{m},\,\varepsilon _{i}$ and $\varepsilon $ (without a
subscript) are the complex dielectric permittivities of a matrix, inclusion and mixture, respectively. 

\section{Wiener and Hashin-Shtrikman bounds for the effective complex permittivities}
Mixing rules must satisfy rigorous bounds which for complex permittivities of composite of two isotropic components have been generalised by  \citet{bergman1980}, \citet{milton1980} and \citet{aspens1982}. It is necessary to discuss three cases of bounds:
\begin{enumerate}
\renewcommand{\theenumi}{\arabic{enumi}.}
\item If we do not know the volume fraction of components and the micro-structure then the resulting permittivity of a mixture is located on a complex surface limited by Wiener bounds:
\begin{enumerate}
\item when there is no screening (all borders of inclusions are parallel to the external electric field)
\begin{equation}
\varepsilon_{p} =f\varepsilon _{a}+\left( 1-f\right)\varepsilon _{b}
\end{equation}

\begin{equation}
\varepsilon_{p} =f\varepsilon _{i}+\left( 1-f\right)\varepsilon _{m}
\end{equation}
\item with maximum screening (all inclusion borders are perpendicular to the external electric field):
\begin{equation}
\frac{1}{\varepsilon_{s}} =\frac{f}{\varepsilon _{a}}+\frac{\left( 1-f\right)}{\varepsilon _{b}}
\end{equation}

\begin{equation}
\frac{1}{\varepsilon_{s}} =\frac{f}{\varepsilon _{i}}+\frac{\left( 1-f\right)}{\varepsilon _{m}}
\end{equation}
\end {enumerate}

\item If we know the volume fraction $f$ of the mixture components and their permittivities $\epsilon_{a}$ and $\epsilon_{b}$ the resulting complex permittivity is located in a smaller area $\Omega'$. The area $\Omega'$ is defined by arcs of circles crossing the three points of $\epsilon_{p}(f)$, $\epsilon_{s}(f)$ and $\epsilon_{a}$ or $\epsilon_{b}$. The permittivities of the mixture components $\epsilon_{a}$ and $\epsilon_{b}$ may have the following values:
\\
$\epsilon_{a}=\epsilon_{i}$, $\epsilon_{b}=\epsilon_{m}$ or $\epsilon_{a}=\epsilon_{m}$, $\epsilon_{b}=\epsilon_{i}$,\\
where $\epsilon_{i}$ and $\epsilon_{m}$ are permittivities of inclusion and matrix, respectively.
The above rigorous bounds are called Hashin-Shtrikman bounds.

\item If the micro-structure is known it is possible to further limit the area in which the resulting permittivity of a mixture must be located.
\end{enumerate}

In this paper the case (2.) has been considered because the volume fraction of inclusions is known but the information about the micro-structure is not available for some mixing rules.

\begin{figure}
\includegraphics[width=84mm]{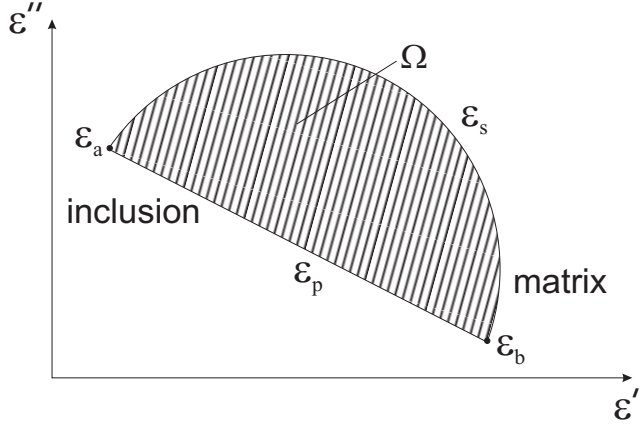} 
\caption{The area where the resulting permittivity is located for case 1.}
\label{figure1}
\end{figure}

\begin{figure}
\includegraphics[width=84mm]{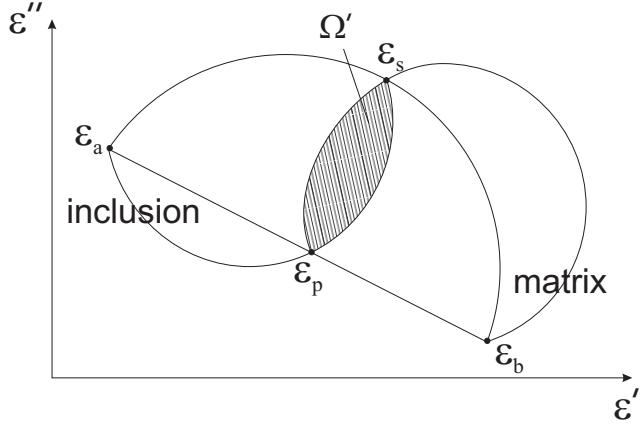} 
\caption{The area where the resulting permittivity is located for case 2.}
\label{figure2}
\end{figure}

\section{Criteria of DDA application}
The criteria for application of DDA have been described in details by \citet{draine1988}. In general the criteria are as follows:
\begin{enumerate}
\item The influence of surface granularity. 

The number of dipoles N must satisfy that 
\begin{equation}
N>N_{min1}\approx 60 \vert m-1 \vert^{3} (\frac{\Delta}{0.1})^{-3},
\label{crit1}
\end{equation}
where $\Delta=0.1$ is the fractional error.

\item Skin depth.
\begin{equation}
N>N_{min2}=\frac{4\pi}{3}\vert m \vert^{3}(\frac{\Delta}{0.1})^{-3}.
\label{crit2}
\end{equation}

\item The influence of magnetic dipole effects. 
\begin{equation}
N>N_{min(magn)}\approx\frac{(ka_{eq})^{3}}{(90\cdot\Delta)^{3/2}}\vert m \vert^{6}=[\frac{(ka_{eq})}{\sqrt{90\cdot\Delta}}]^{3}\vert m \vert^{6},
\label{crit3}
\end{equation}
where ${\rm k=\frac{2\pi}{\lambda}}$ and ${\rm a_{eq}=0.15 \mu m}$. 
The above equation combined with the criterion of influence of skin depth gives: 
\begin{equation}
N>N_{min2}\approx \frac{4\pi}{3}(ka_{eq})^{3}\vert m \vert^{3}(\frac{\Delta}{0.1})^{-3}[1+\frac{\vert m \vert^{3}}{36\pi}\frac{\Delta}{0.1})^{3/2}]
\label{crit4}
\end{equation}

\end{enumerate}
The criteria given by \citet{draine1988} (Equations~\ref{crit1} and \ref{crit4}) allow to determine the conditions which must be satisfied in order to use the DDA method. One may either calculate the smallest number of dipoles at a given wavelength range or determine the wavelength range for a given number of dipoles. In our case the criteria have been used to determine the range of wavelengths for the considered mixtures in order to calculate the ${\rm Q_{ext}}$ using the DDA. For this purpose two mixture cases described by equations (6) and (8) have been used.

\begin{figure}
\includegraphics[angle=270,width=84mm]{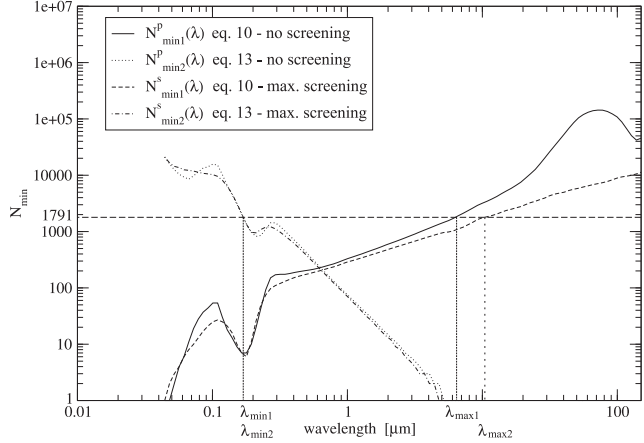} 
\caption{An illustrative example showing the choice of the wavelength range which is in accordance with Draine criteria. See text.}
\label{criteria}
\end{figure}

For the calculated values of ${\rm \epsilon_{p}}$ and ${\rm \epsilon_{s}}$ for the given volume fraction of inclusions $f$ in the  wavelength range from $0.0443$ to $\rm 150 \mu m$  the minimum number of dipoles $N_{min1}^{p}(\lambda)$ and $N_{min1}^{s}(\lambda)$ have been calculated from (\ref{crit1}) and $N_{min2}^{p}(\lambda)$  and $N_{min2}^{s}(\lambda)$ from (\ref{crit4}). Figure \ref{criteria} shows an example of the relations of those values for 30\% inlusions of graphite in amorphous carbon matrix. In the wavelength range for which the following ralations are simultaneously satisfied the Draine criteria are also satified:
\begin{itemize}
\item $N_{min1}^{p}(\lambda)<1791$
\item $N_{min1}^{s}(\lambda)<1791$
\item $N_{min2}^{p}(\lambda)<1791$  
\item $N_{min2}^{s}(\lambda)<1791$  
\end{itemize}

The two chosen values have been the limits of DDA applicability for the given mixture in terms of composition and volume fraction of inclusions $f$. The choice of wavelength ranges have been applied to ${\rm f=0.05 - 0.50}$ with step of $0.05$.  The value of $f$ has been limited to 0.5 because it is the approximate value of percolation threshold. The same procedure was carried out for all mixtures although after reaching the value $f=0.5$ the whole procedure has been carried with the roles of matrix and inclusions interchanged.

\section{Details of calculations}
\begin{figure}
\includegraphics[width=164mm]{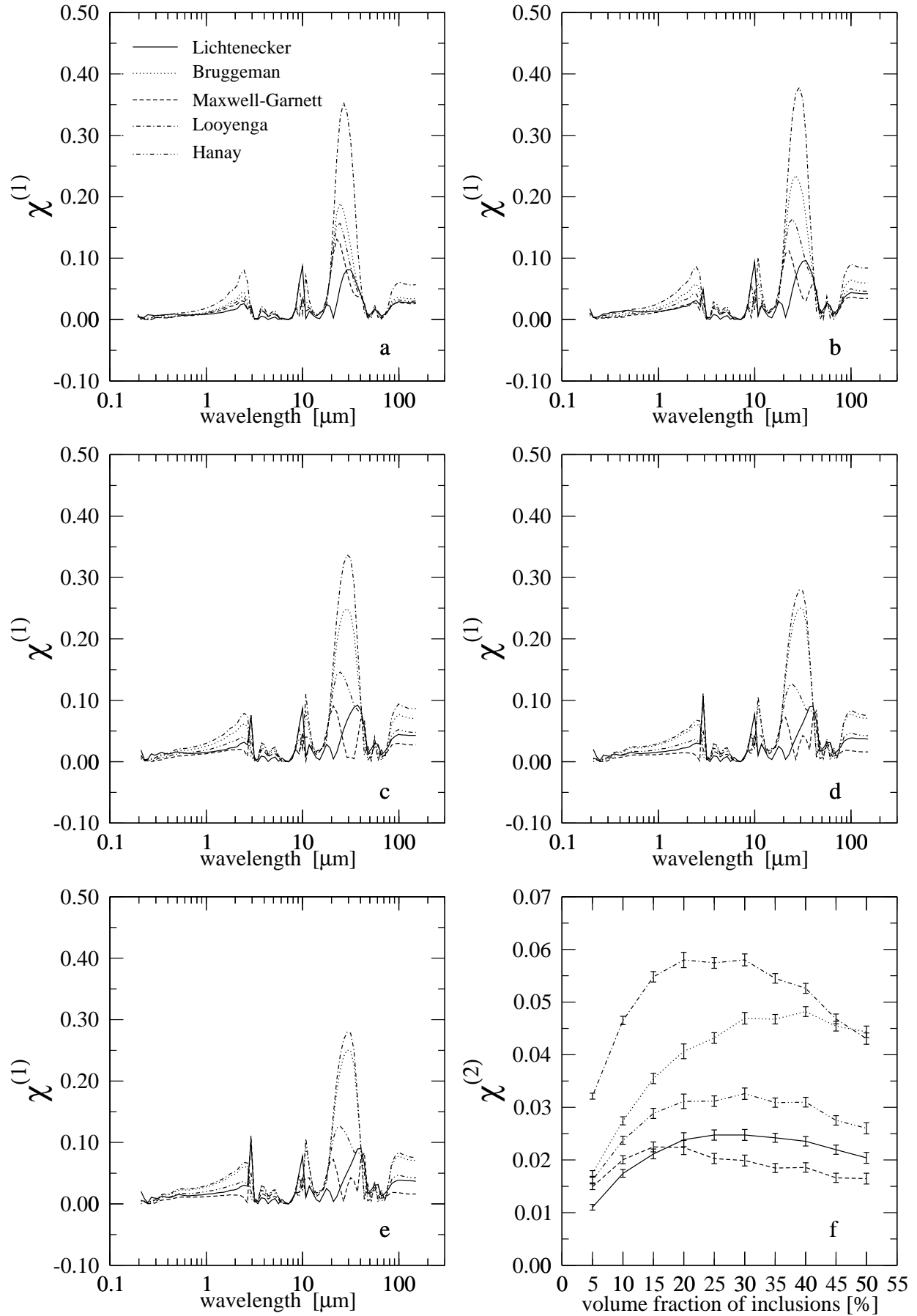} 
\caption{Ice - matrix, silicate - inclusions (a-e volume fractions of inclusions from 10\% to 50\% with 10\% step, f - best fit in the whole range of wavelengths depending on volume fraction of inclusions)}
\label{l_sil}
\end{figure}

\begin{figure}
\includegraphics[width=164mm]{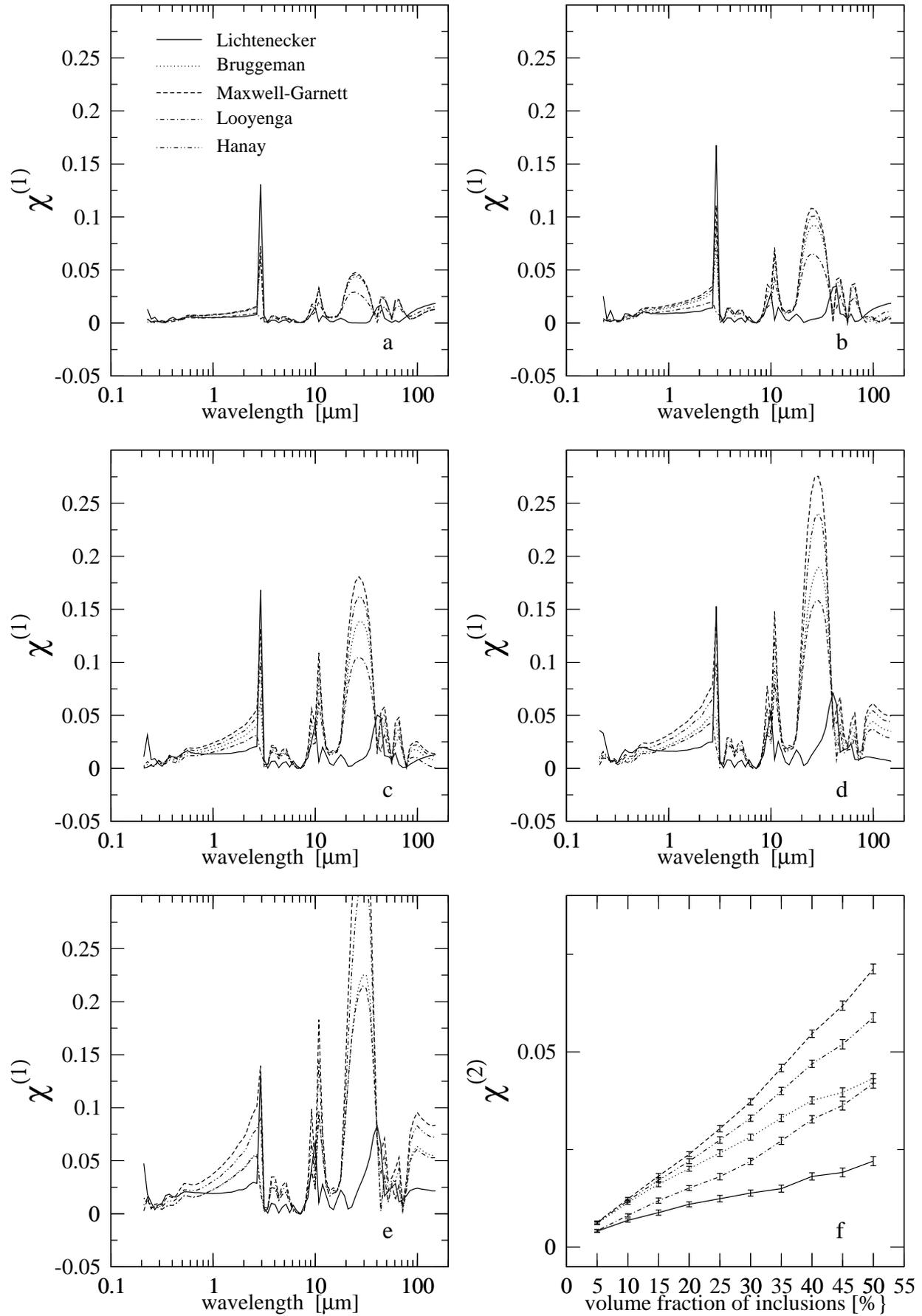} 
\caption{Silicate - matrix, ice - inclusions (same as in Fig.~\ref{l_sil})}
\label{sil_l}
\end{figure}

\begin{figure}
\includegraphics[width=164mm]{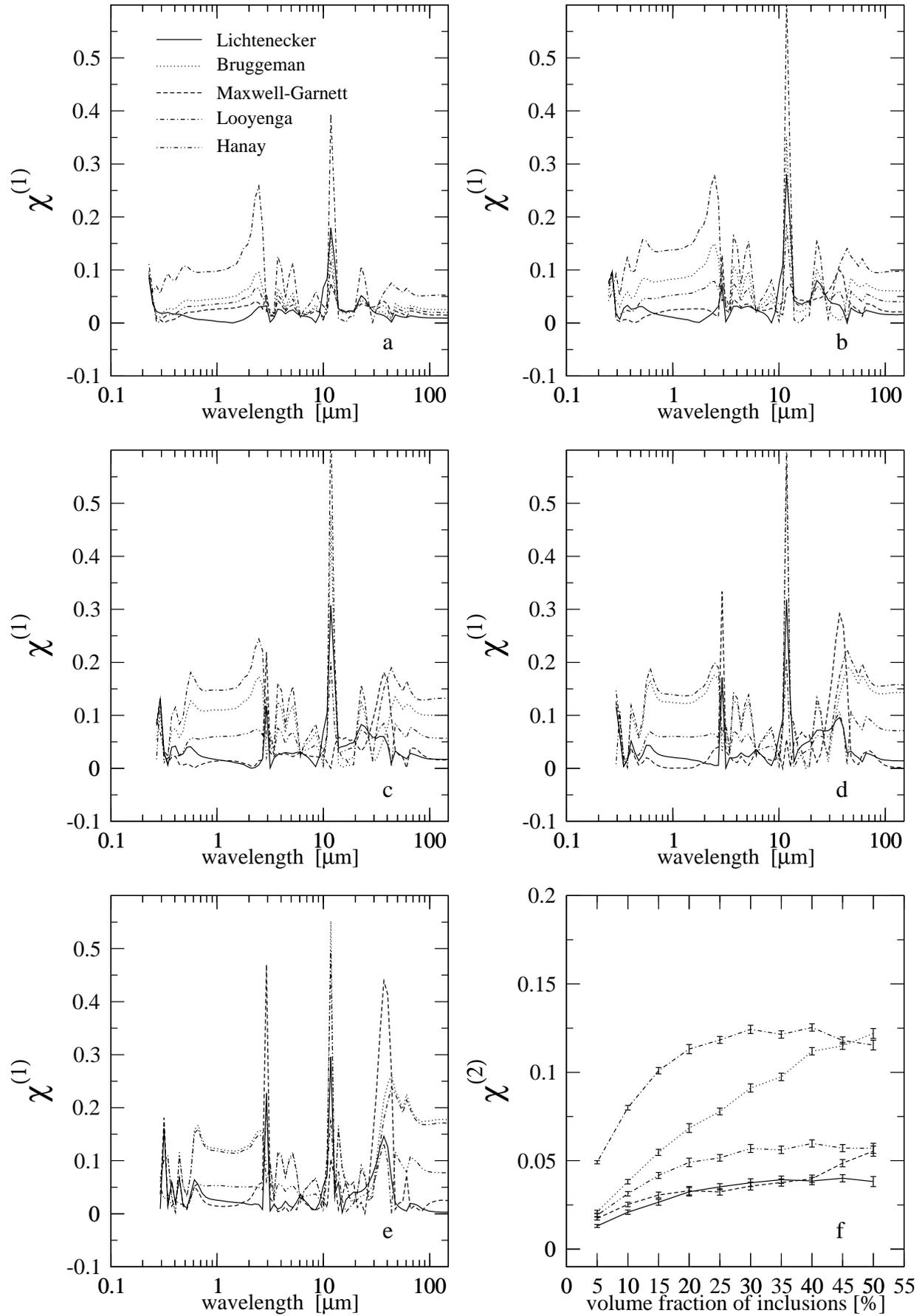} 
\caption{Ice - matrix, SiC - inclusions (same as in Fig.~\ref{l_sil})}
\label{l_sic}
\end{figure}

\begin{figure}
\includegraphics[width=164mm]{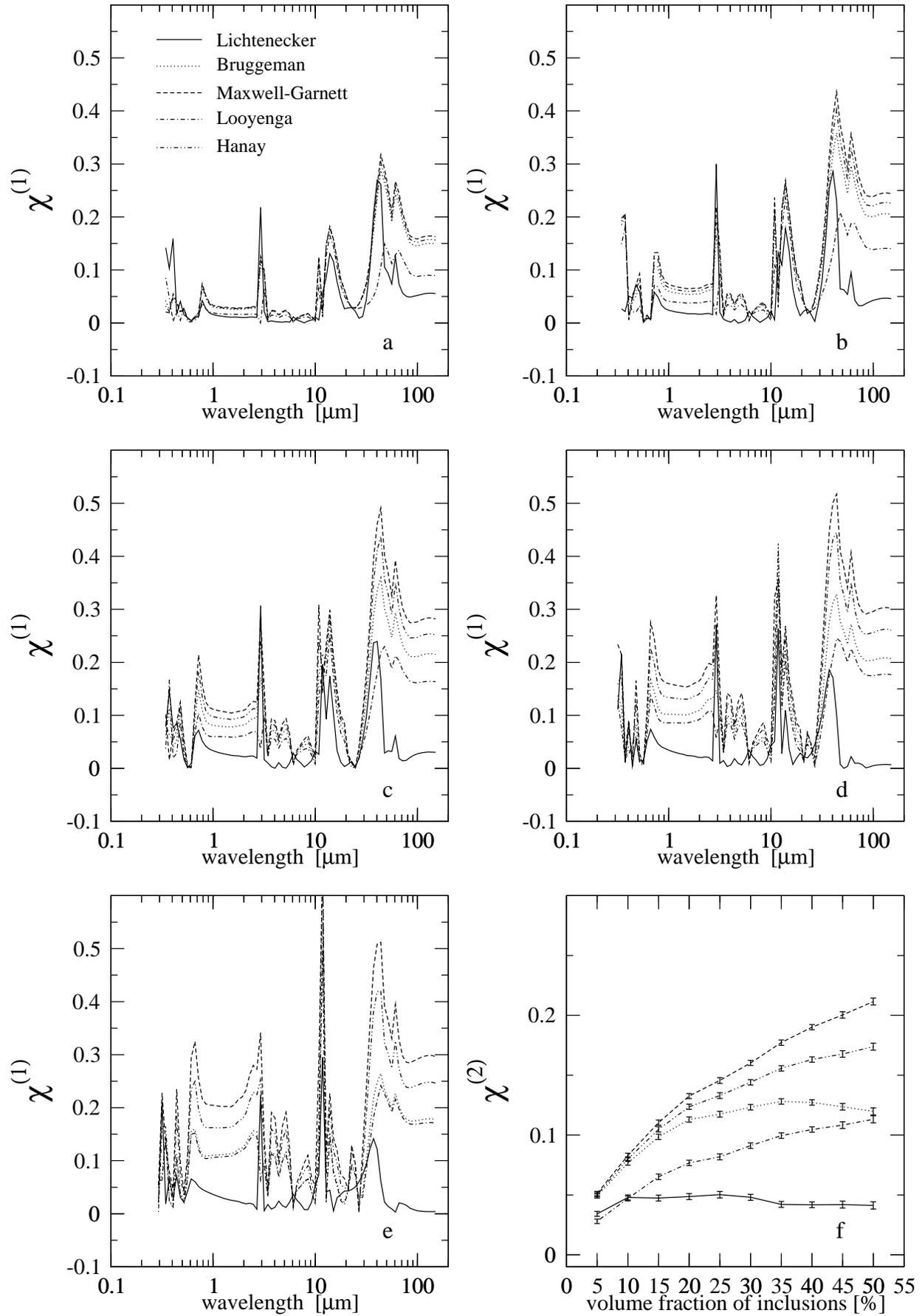} 
\caption{SiC - matrix, ice - inclusions (same as in Fig.~\ref{l_sil})}
\label{sic_l}
\end{figure}

\begin{figure}
\includegraphics[width=164mm]{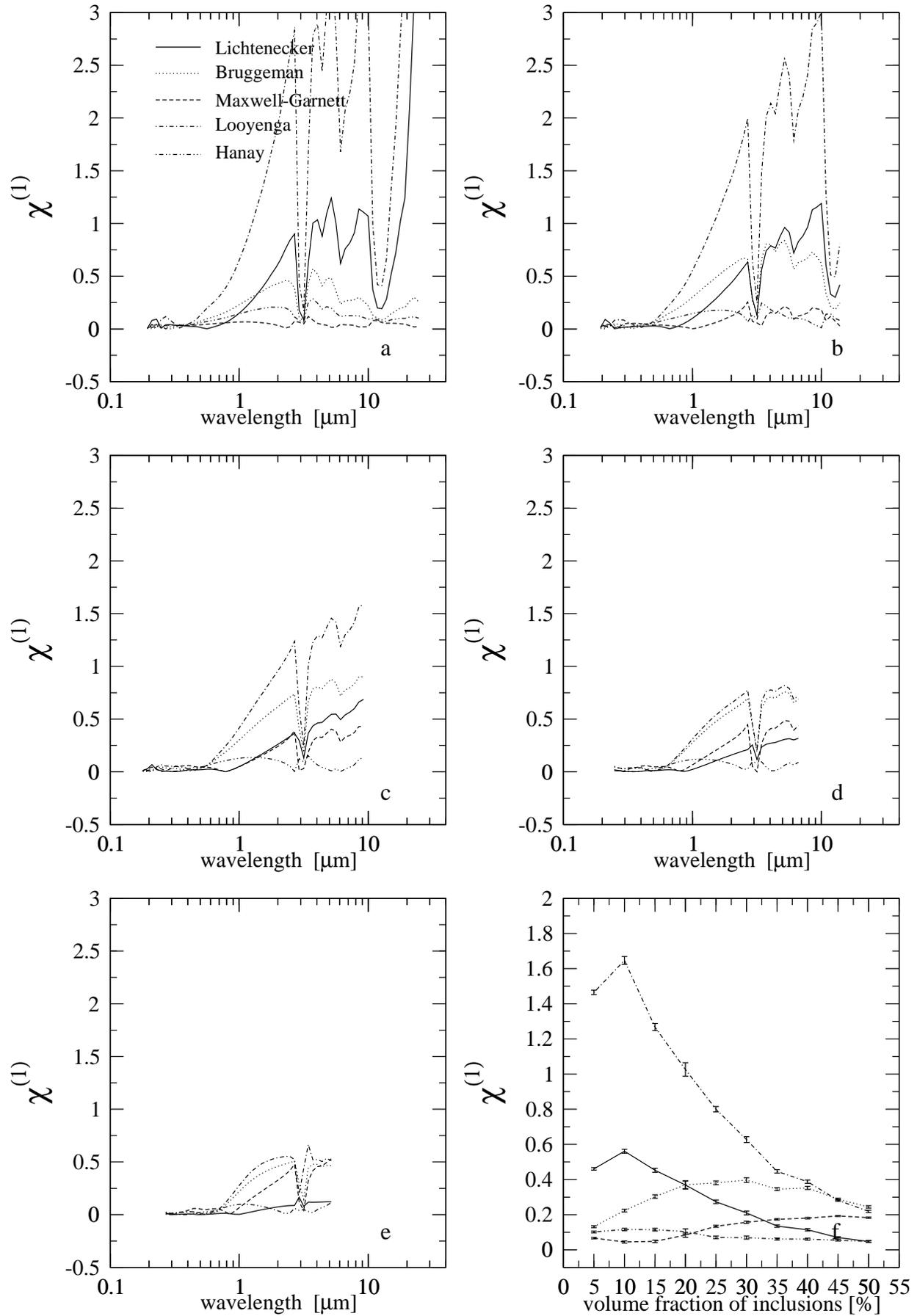} 
\caption{Ice - matrix, graphite - inclusions (same as in Fig.~\ref{l_sil})}
\label{l_g}
\end{figure}

\begin{figure}
\includegraphics[width=164mm]{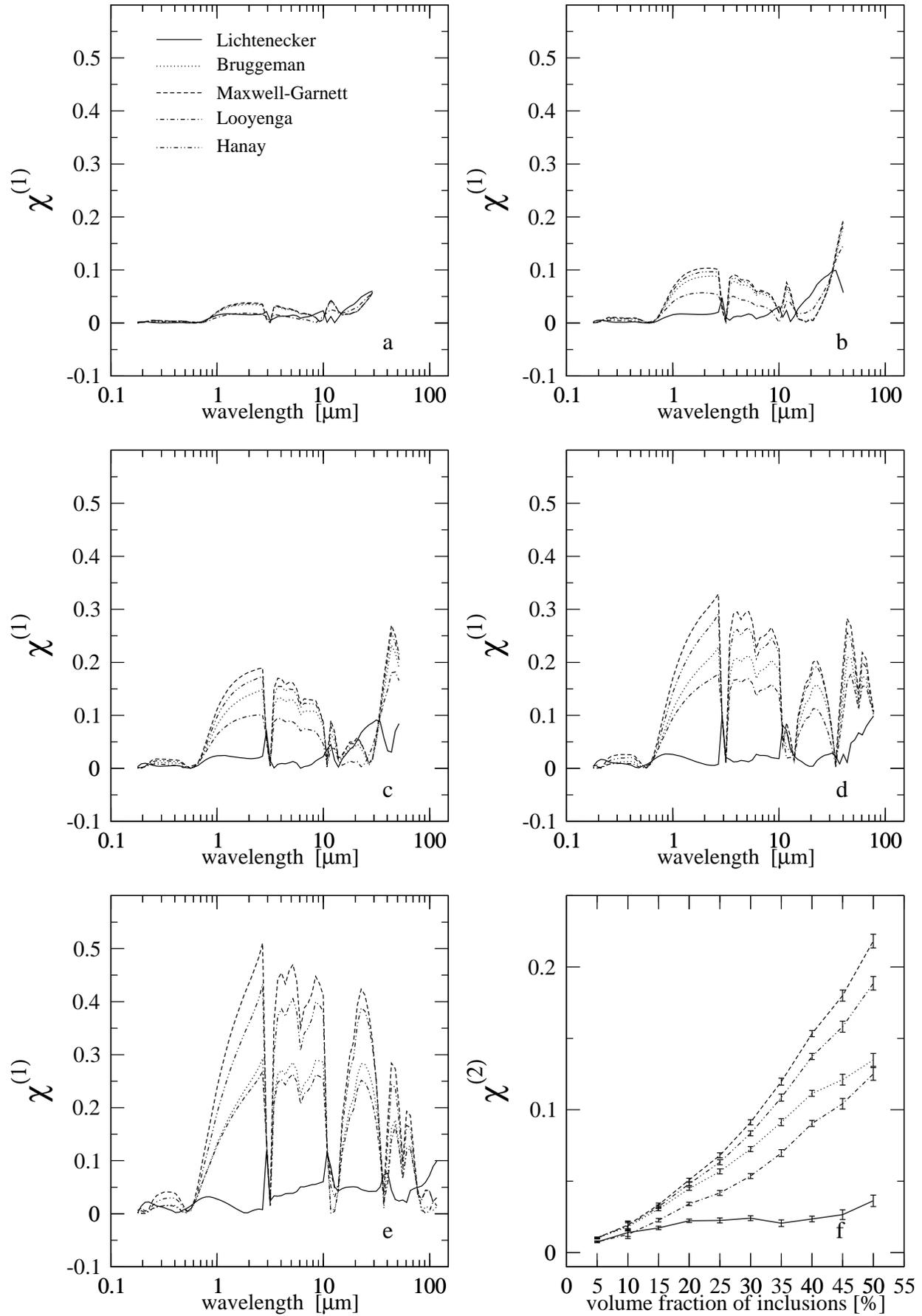} 
\caption{Carbon - matrix, ice - inclusions (same as in Fig.~\ref{l_sil})}
\label{c_l}
\end{figure}

\begin{figure}
\includegraphics[width=164mm]{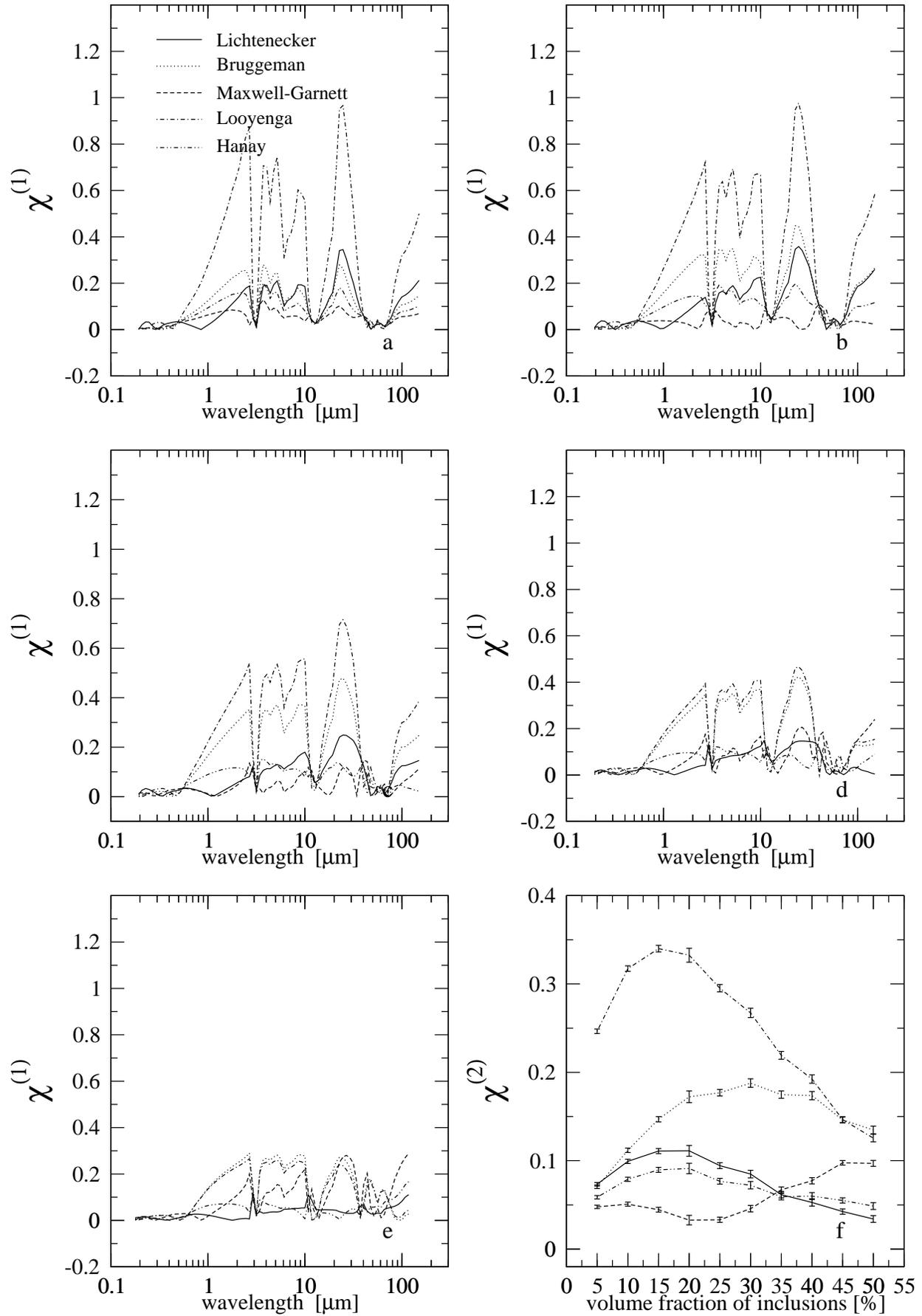} 
\caption{Ice - matrix, carbon - inclusions (same as in Fig.~\ref{l_sil})}
\label{l_c}
\end{figure}

\begin{figure}
\includegraphics[width=164mm]{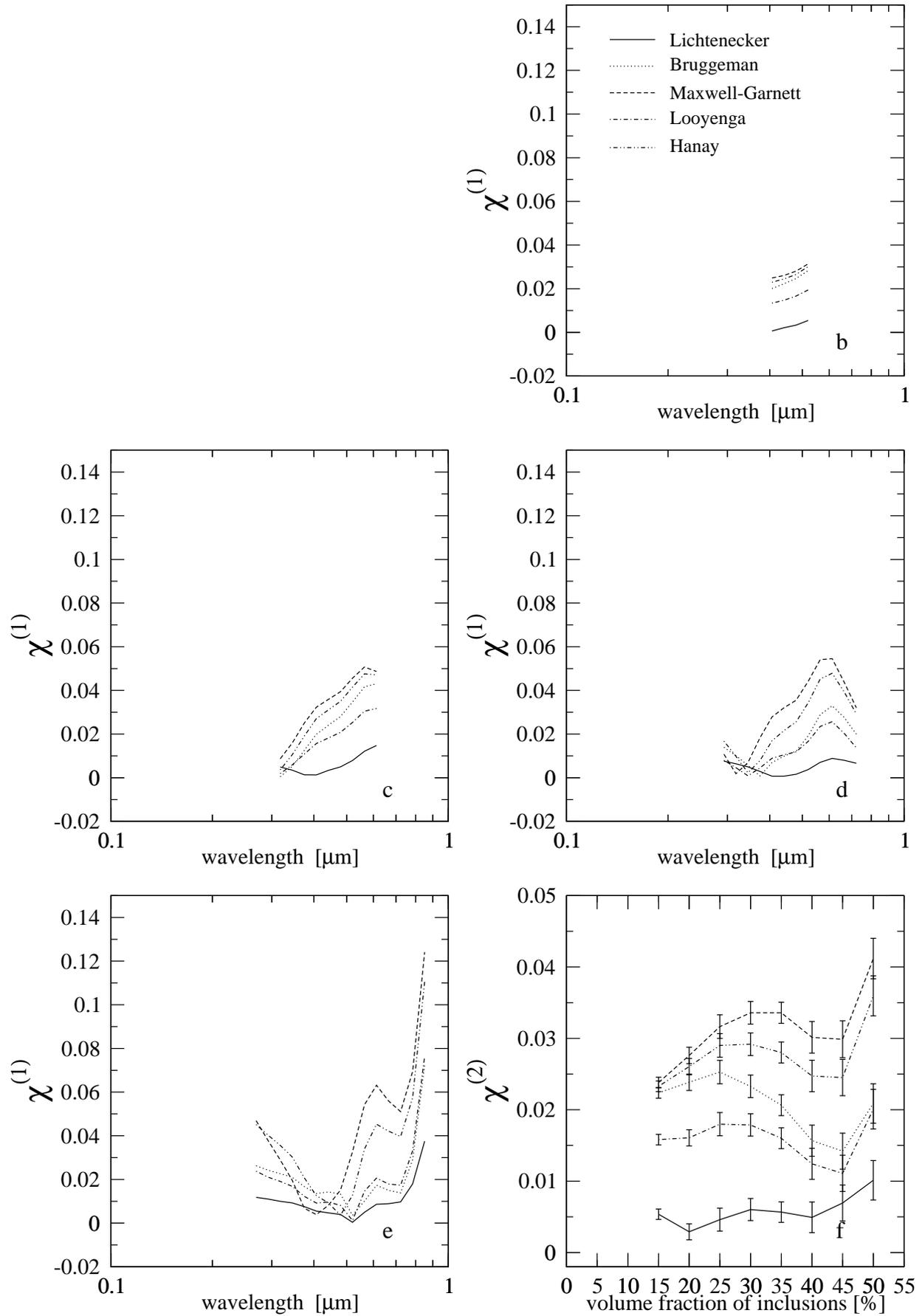} 
\caption{Fe - matrix, ice - inclusions (same as in Fig.~\ref{l_sil})}
\label{fe_l}
\end{figure}

\begin{figure}
\includegraphics[width=164mm]{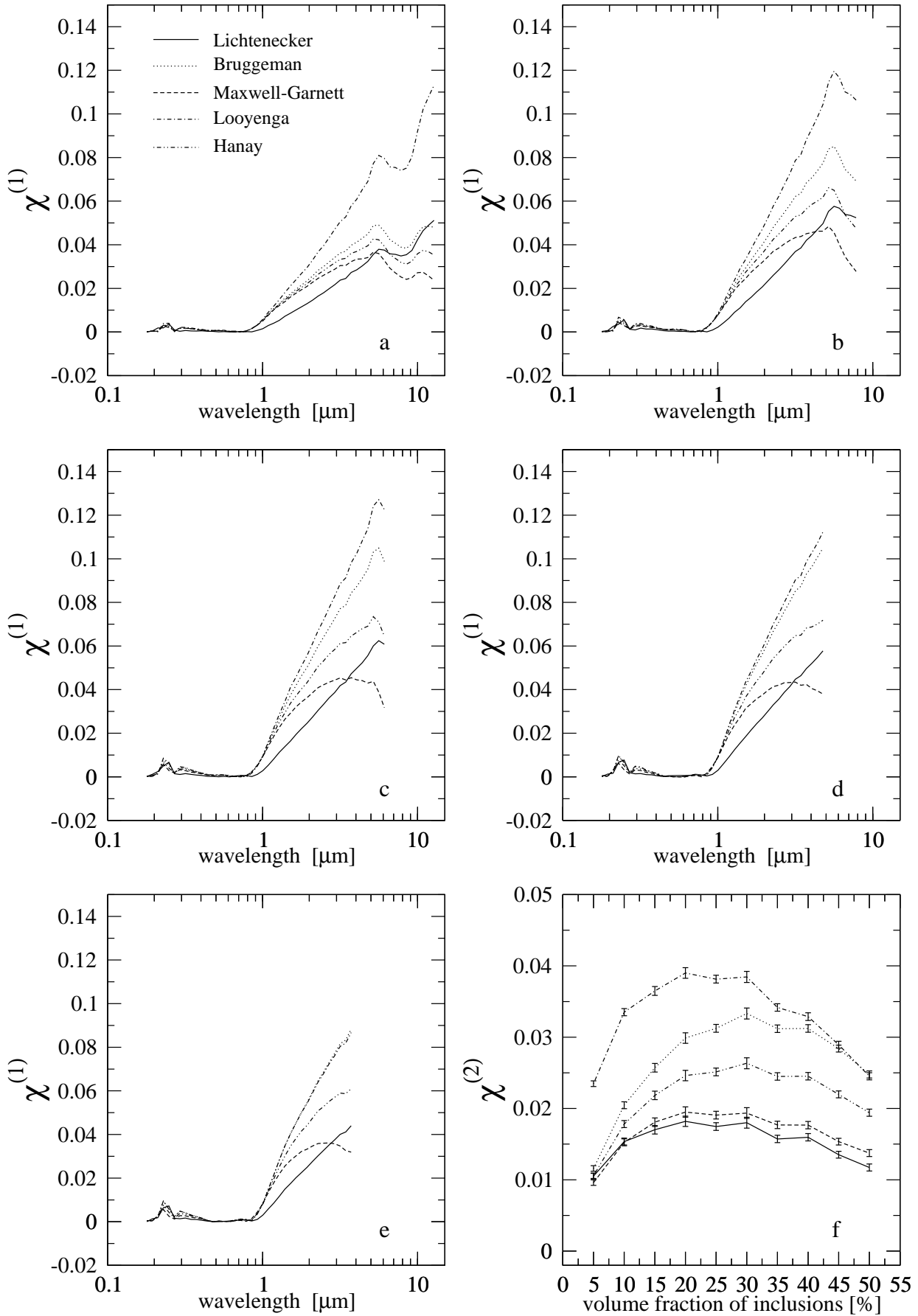} 
\caption{Carbon - matrix, graphite - inclusions (same as in Fig.~\ref{l_sil})}
\label{c_g}
\end{figure}

\begin{figure}
\includegraphics[width=164mm]{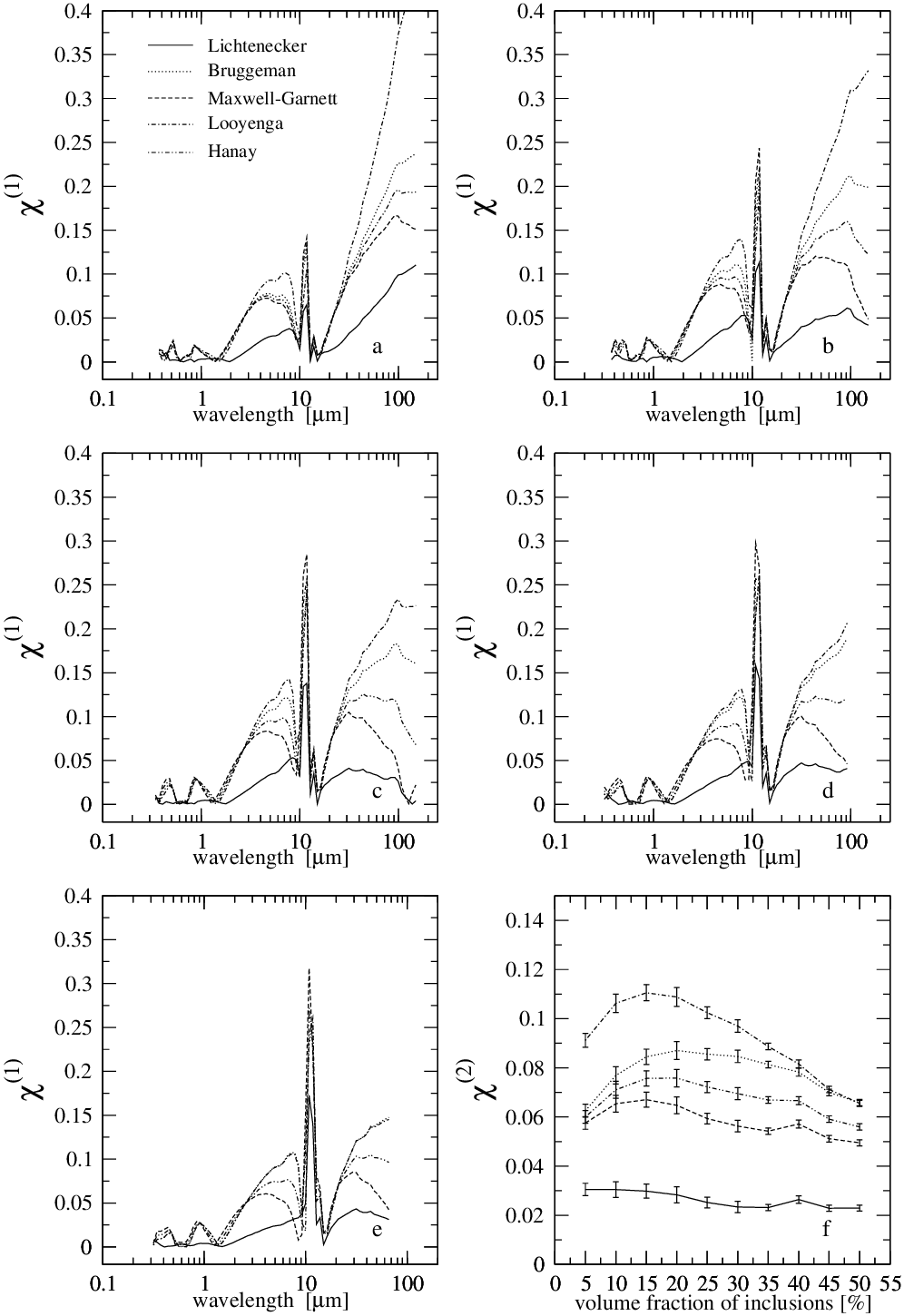} 
\caption{SiC - matrix, carbon - inclusions (same as in Fig.~\ref{l_sil})}
\label{sic_c}
\end{figure}

\begin{figure}
\includegraphics[width=164mm]{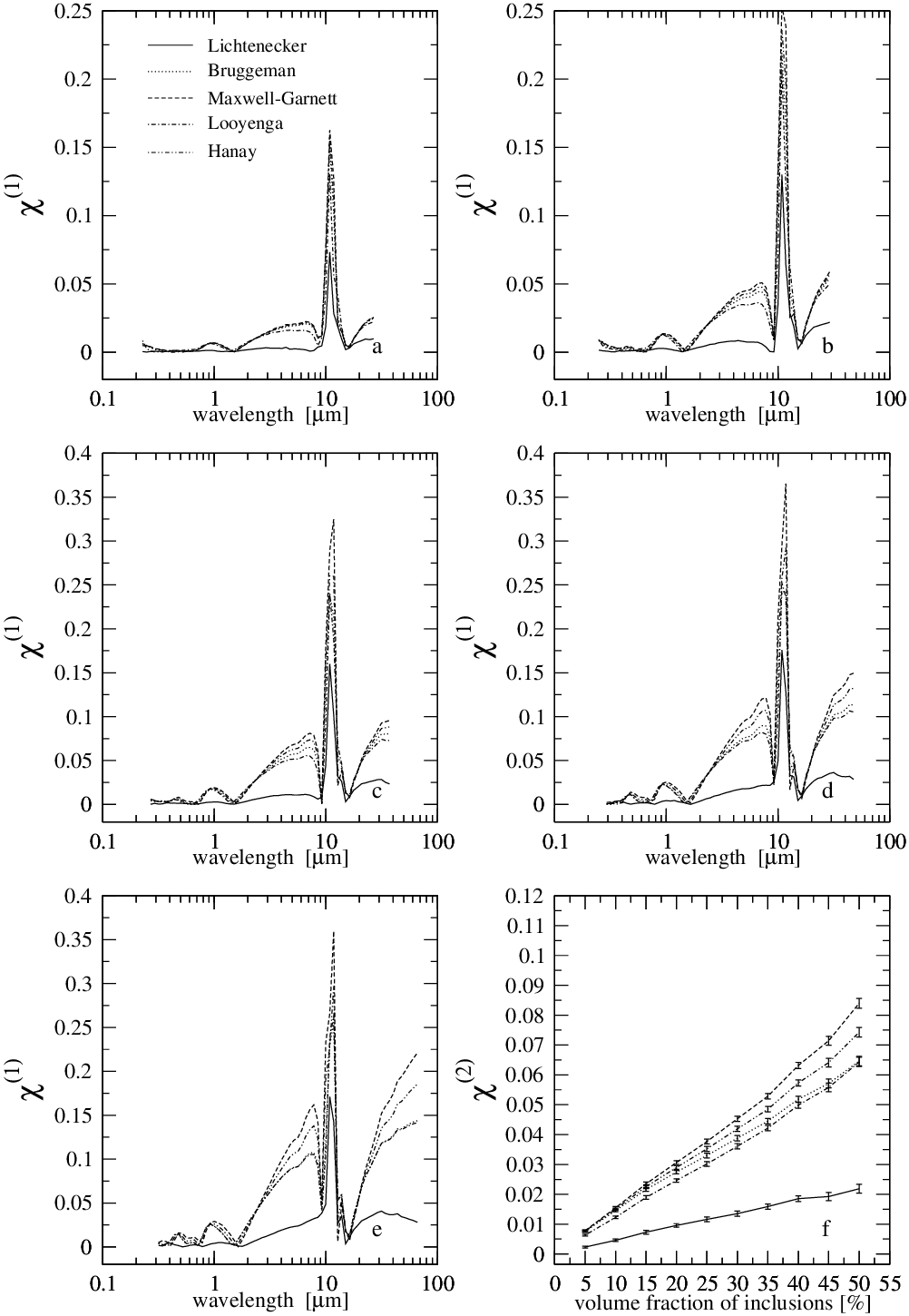} 
\caption{Carbon - matrix, SiC - inclusions (same as in Fig.~\ref{l_sil})}
\label{c_sic}
\end{figure}

\begin{figure}
\includegraphics[width=164mm]{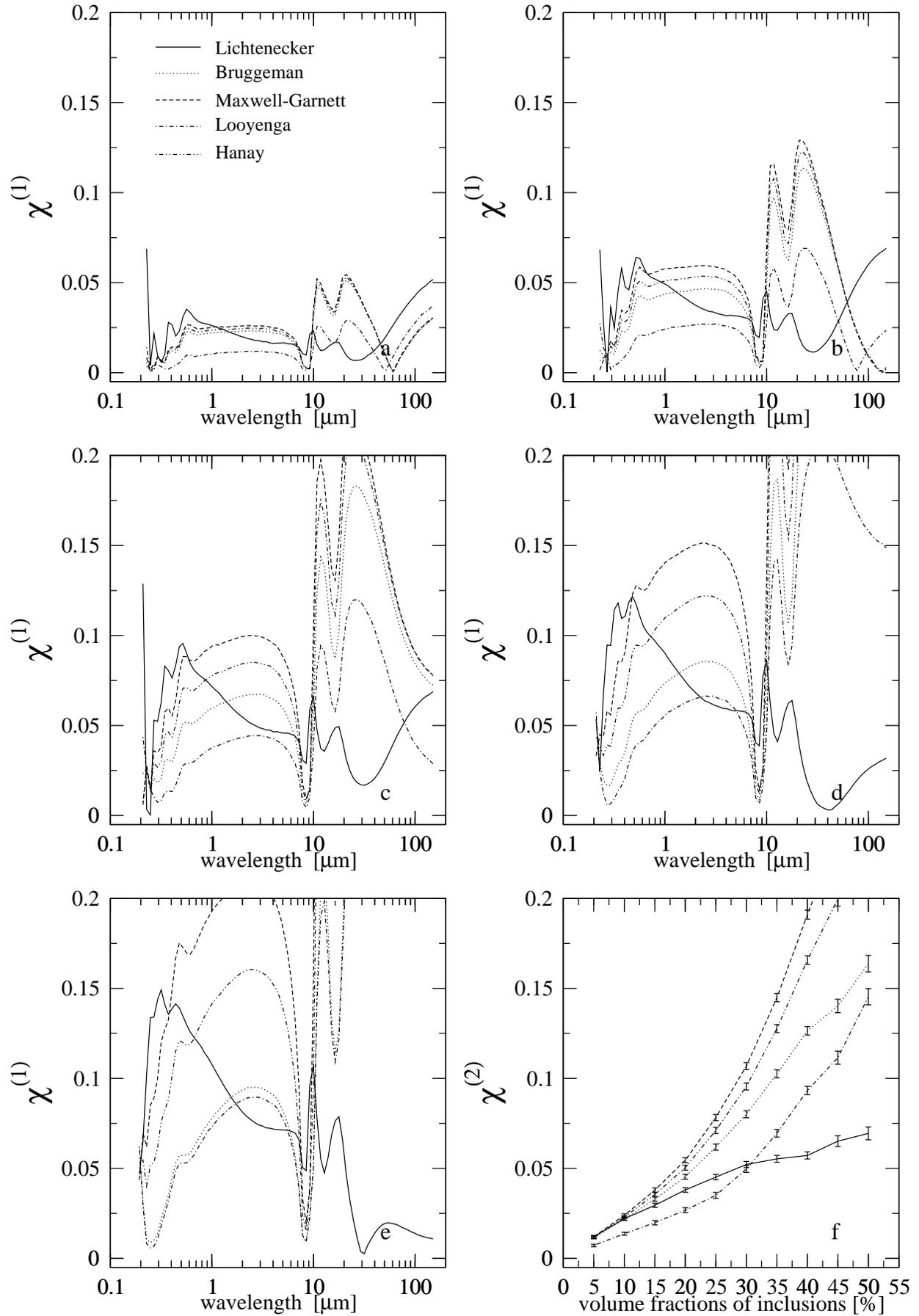} 
\caption{Silicate - matrix, vacuum - inclusions (same as in Fig.~\ref{l_sil})}
\label{sil_vac}
\end{figure}

\begin{figure}
\includegraphics[width=164mm]{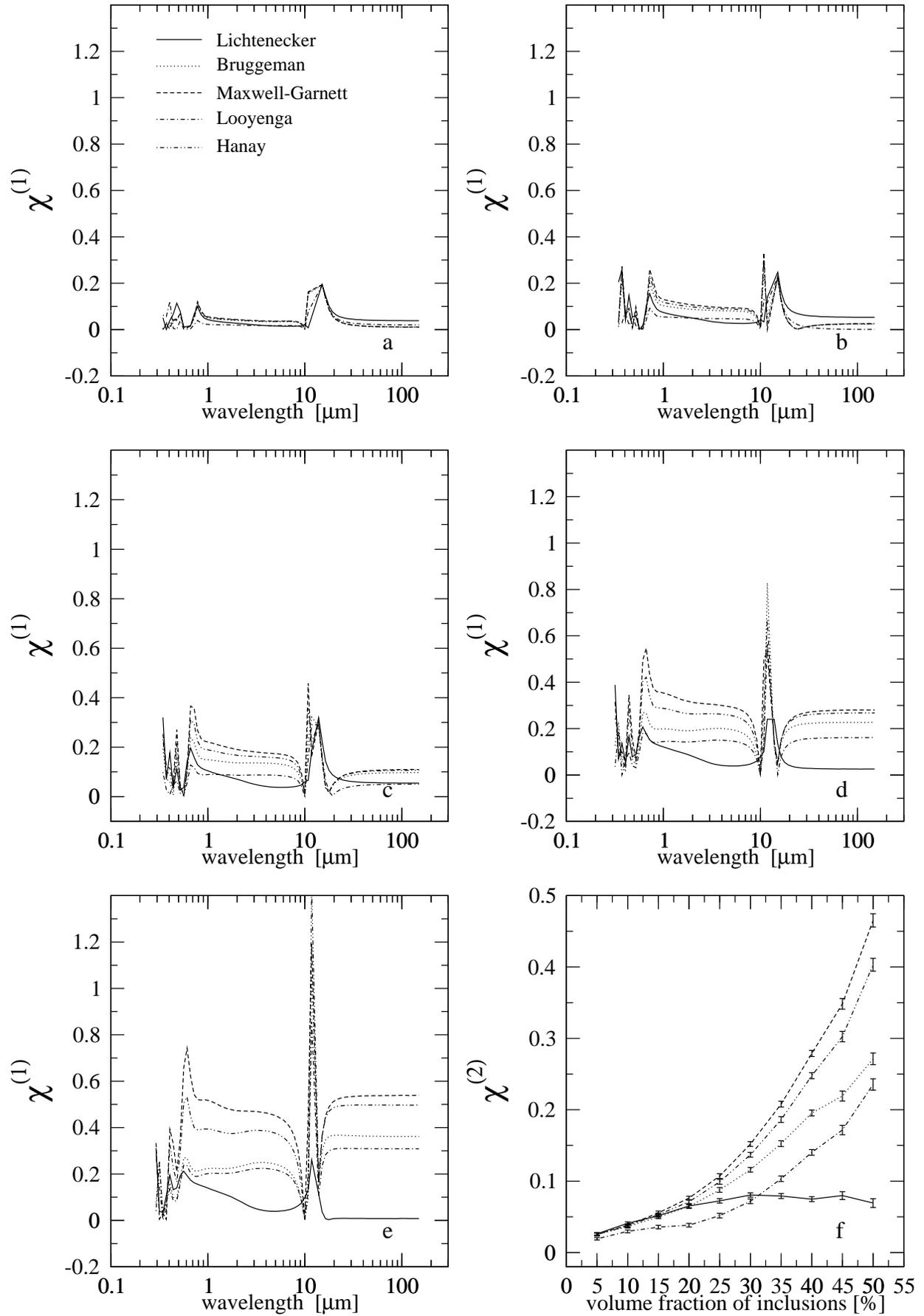} 
\caption{SiC - matrix, vacuum - inclusions (same as in Fig.~\ref{l_sil})}
\label{sic_vac}
\end{figure}

\begin{figure}
\includegraphics[width=164mm]{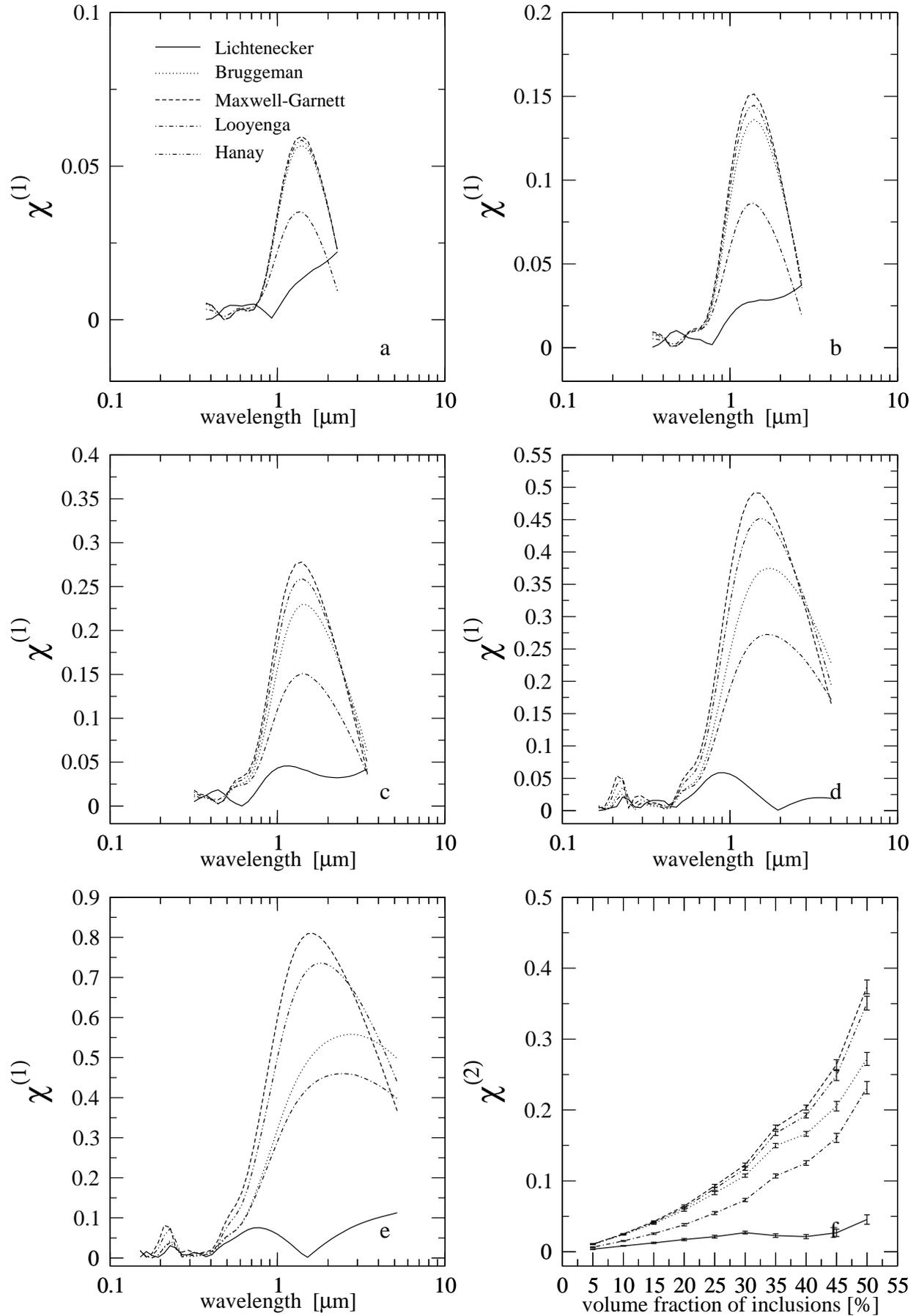} 
\caption{Graphite - matrix, vacuum - inclusions (same as in Fig.~\ref{l_sil})}
\label{g_vac}
\end{figure}

\begin{figure}
\includegraphics[width=164mm]{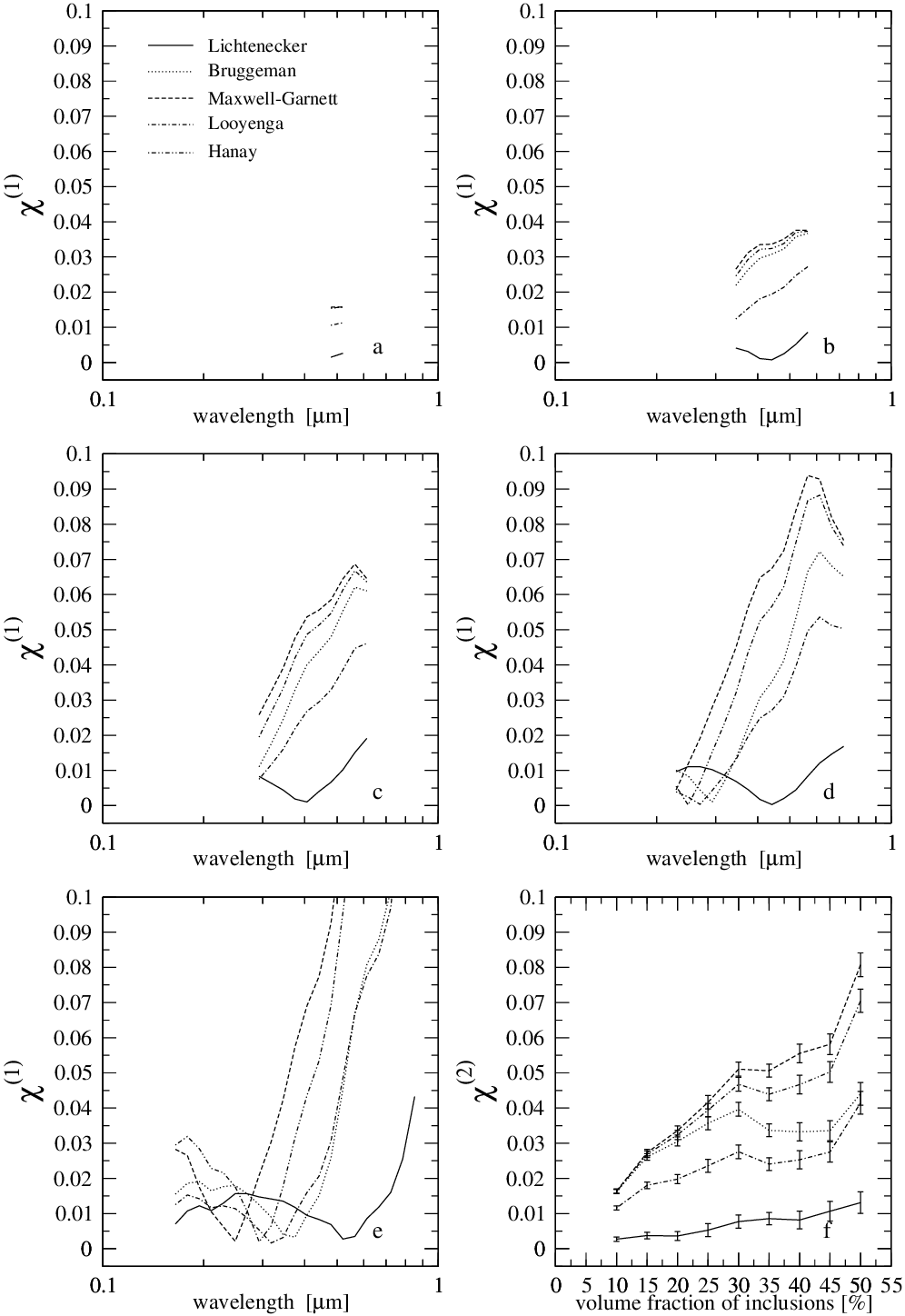} 
\caption{Fe - matrix, vacuum - inclusions (same as in Fig.~\ref{l_sil})}
\label{fe_vac}
\end{figure}

\begin{figure}
\includegraphics[width=164mm]{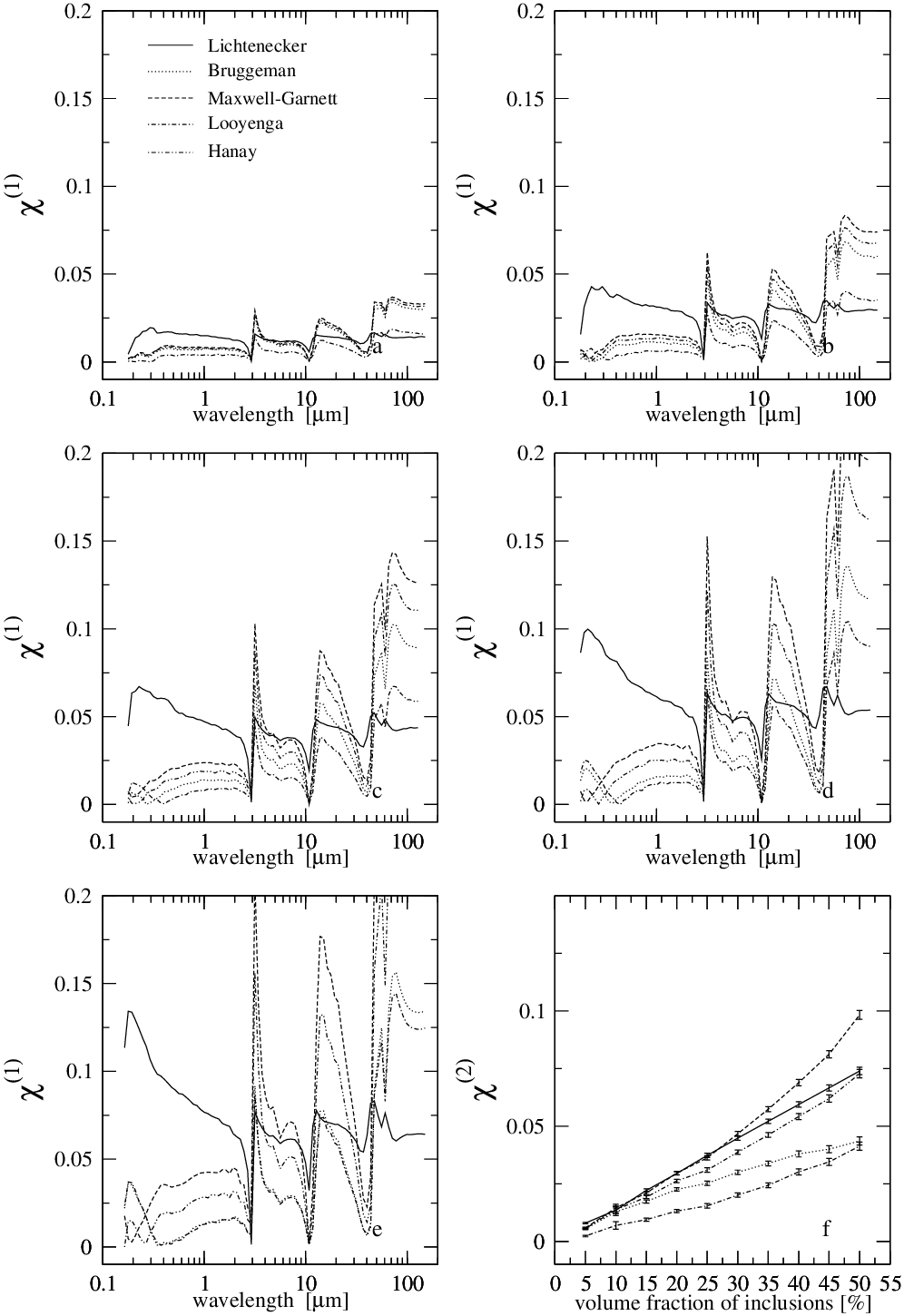} 
\caption{Ice - matrix, vacuum - inclusions (same as in Fig.~\ref{l_sil})}
\label{ice_vac}
\end{figure}

The details of calculation have been given in the previous paper \citep{maron2005} with the only difference that we have chosen the mixtures of amorphous carbon, graphite, SiC, "astronomical silicate" and iron in ice and mixtures of ice in those materials for examination. Besides, we have considered each material with pores containing vacuum. The calculations have been carried out for volume fractions from 5\% to 50\% with step of 5\%. Draine has published a new version of DDSCAT program but there were no changes in the domain that was interesting for this work. Therefore in current work as in the previous we have used the version DDSCAT 5a10. The values of permittivity for mixtures calculated according to equations (1)-(5) have been chosen with respect to the Wiener and Hashin-Shtrikman bounds (it was important for Hanay and Bruggeman rules where there were 3 or 2 solutions of mixing equations, respectively, and also in case of porous ice). The wavelength ranges for each mixture have been bound according to Draine criteria which is seen in Figures~\ref{l_sil}-\ref{ice_vac}. Many authors (including \citet{draine1988} and \citet{wolff1994}) have examined the influence of the number of dipoles on physical convergence of extinction coefficient. Numerical tests indicate that when the number of dipoles in the DDA method aproach infinity the extinction coefficient for the pseudosphere $\rm Q_{ext}^{DDA}$ aproaches the extinction coefficient obtained from the Mie theory ($\rm Q_{ext}^{DDA}(N\rightarrow\infty)=Q_{ext}^{Mie}$). Of course, using a large number of dipoles in the grain implies inclusion consisting of many dipoles as well because its radius should not be smaller than $\rm 50\AA$ \citep{bohren1977}. \citet{wolff1998} examining porous grains concluded that extinction obtained by DDA method for single dipole vacuum inclusion was in good accordance with that obtained from Mie theory and effective extinction coefficient calculated from effective medium theory, while for multi dipole vacuum inclusions they were strikingly not compliant with each other. Similar conclusions were made by \citet{voshchin2007} who stated that if the inclusions were not simple dipoles in the DDA terms the scattering charcteristics of aggregates were not well reproduced by the EMT calculations. Therefore, the inclusions have been assumed to be simple dipoles. We have limited the number of dipoles in the grain to a rather small number (1791) for the following reasons:
\begin{enumerate}
\item The size of single dipole inclusions ($\rm r=100\AA$) are sufficient in order to safely use the bulk dielectric function.
\item Tests by \citep{draine1988} showed that if the criteria (eq.~\ref{crit1}) and (eq.~\ref{crit4}) are satisfied we obtain a good agreement between extinction coefficients from DDA with those calculated from Mie theory.
\item In our approach we do not pursue the convergence of DDA results with those obtained from Mie/EMT but  only the single dipole inclusions DDA with DDA/EMT. Therefore, the limited number of dipoles is sufficient providing that the Draine criteria are satisfied.
\end{enumerate}

\begin{figure}
\includegraphics[width=84mm]{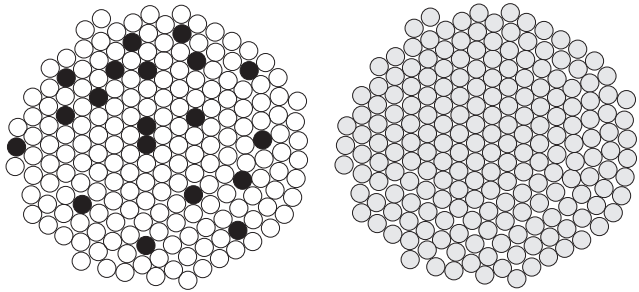} 
\caption{Schematic representation of grain (left - white circles depict matrix dipoles and the black ones dipole inclusions, right - grey circles are uniform dipoles)}
\label{grain}
\end{figure}

As it has been stated by \citet{wolff1994} the inclusions indeed do not have to be dipoles. \citet{wolff1994} were the first to consider nonspherical and nondipole inclusions containing a large number of dipoles. This aproach is justified in case of fitting the theoretical extinction curve obtained from the DDA method with the observed one. Nevertheless, in this paper we compare the extiction curve calculated for grains composed of two kinds of dipoles with the one calculated for grains obtained by using a mixing rule. This situation is illustrated by the figure \ref{grain}. In Fig. \ref{grain}(left) the white circles stand for matrix dipoles ($\rm \epsilon_{m}$) and the black ones for dipole inclusions ($\rm \epsilon_{i}$) of the mixture with 20\% of inclusions. In Fig. \ref{grain}(right) the grey circles stand for uniform dipoles with permittivity obtained from mixing rules ($\rm \epsilon$) for the same volume fraction of inclusions. Both cases have been computed using the DDA. Therefore, it is justified to use the idealised grains and inclusions.

The arrangement of inclusions (DDA elements) is random. In order to obtain random number of DDA elements in the discrete dipole arrays we have used random number generator "Research Randomizer" available at {\it http://www.randomizer.org}. We have generated 10 series of numbers corresponding to the given volume fractions of inclusions out of all 1791 dipoles. The generated numbers have been sorted in ascending order. The location of dipoles in the array for the spherical particle obtained from the routine {\it calltarget.f} was the same as for homogeneous grains with the only difference that for the randomly generated numbers of the DDA elements they had the refractive index of inclusions and the remaining elements had the refractive index of a matrix. Certainly the influence of inclusions topology on extinction might exist but   the authors of the considered mixing rules have assumed a statistical distribution of inclusions. Therefore, in our calculations the random distributions have been used. Because there is a scattering of results for different random distributions the efficiency factors for extinction have been calculated for 10 different distributions and then averaged. For grains with radii of ${\rm r=0.15 \mu m}$ 10 values of efficiency factors for extinction ${\rm Q_{i,j,l}^{rand}}$ depending on random location have been calculated. The calculations using the computer program DDSCAT.5a10 have been carried out for wavelengths in the range permitted by the Draine criteria for the given refractive indices. The subscript i in the symbol ${\rm Q_{i,j,l}^{rand}}$ corresponds to the number of random location of inclusions, j - the volume fraction of inclusion and the subscript l corresponds to wavelength. Next the mean extinction has been calculated as: 
\begin{equation}
Q_{l,j}^{rand}=\frac{1}{10}\sum_{i=1}^{10}Q_{i,j,l}^{rand}.
\end{equation}

We have calculated the standard deviation of the mean:
\begin{equation}
\sigma_{l,j}=\sqrt{\frac{\sum_{i=1}^{I}(Q_{l,j}^{rand}-Q_{i,j,l}^{rand})^2}{I(I-1)}},
\label{error1}
\end{equation}
where I=10 is the number of random locations.

The extinction for homogeneous grains has been calculated using DDA assuming that all elements (dipoles) consist of the same mixture with averaged refractive index calculated from the given mixing rule. In the obtained extinction coefficient for the homogeneous grain ${\rm Q_{l,j,p}^{homog}}$ the subscripts l and j denote the wavelength and volume fraction of inclusion respectively, and the subscript p denotes the given mixing rule. The relative deviation $\chi_{l,j,p}^{(1)}$ (further used as $\chi^{(1)}$) was calculated from

\begin{equation}
\chi_{l,j,p}^{(1)}=\frac{\left| Q_{l, j}^{rand}-Q_{l,j,p}^{homog} \right|}{Q_{l, j}^{rand}},
\end{equation}
where  $Q_{l, j}^{rand}$ is averaged extinction coefficient for randomly located inclusions, $Q_{l,j,p}^{homog}$ is extinction coefficient for homogeneous grains for a given mixing rule.

The values of $\chi_{l,j,p}^{(1)}$ are biased by deviations related to Equation \ref{error1} in the following way:
\begin{equation}
\Delta\chi_{l,j,p}^{(1)}=\left| \frac{\partial \chi_{l,j,p}^{(1)}}{\partial Q_{l,j}^{rand}}\right| \sigma_{l,j}^{rand}=Q_{l,j,p}^{homog}(Q_{l,j}^{rand})^{-2} \sigma_{l,j}
\end{equation}

\begin{figure}
\includegraphics[width=84mm]{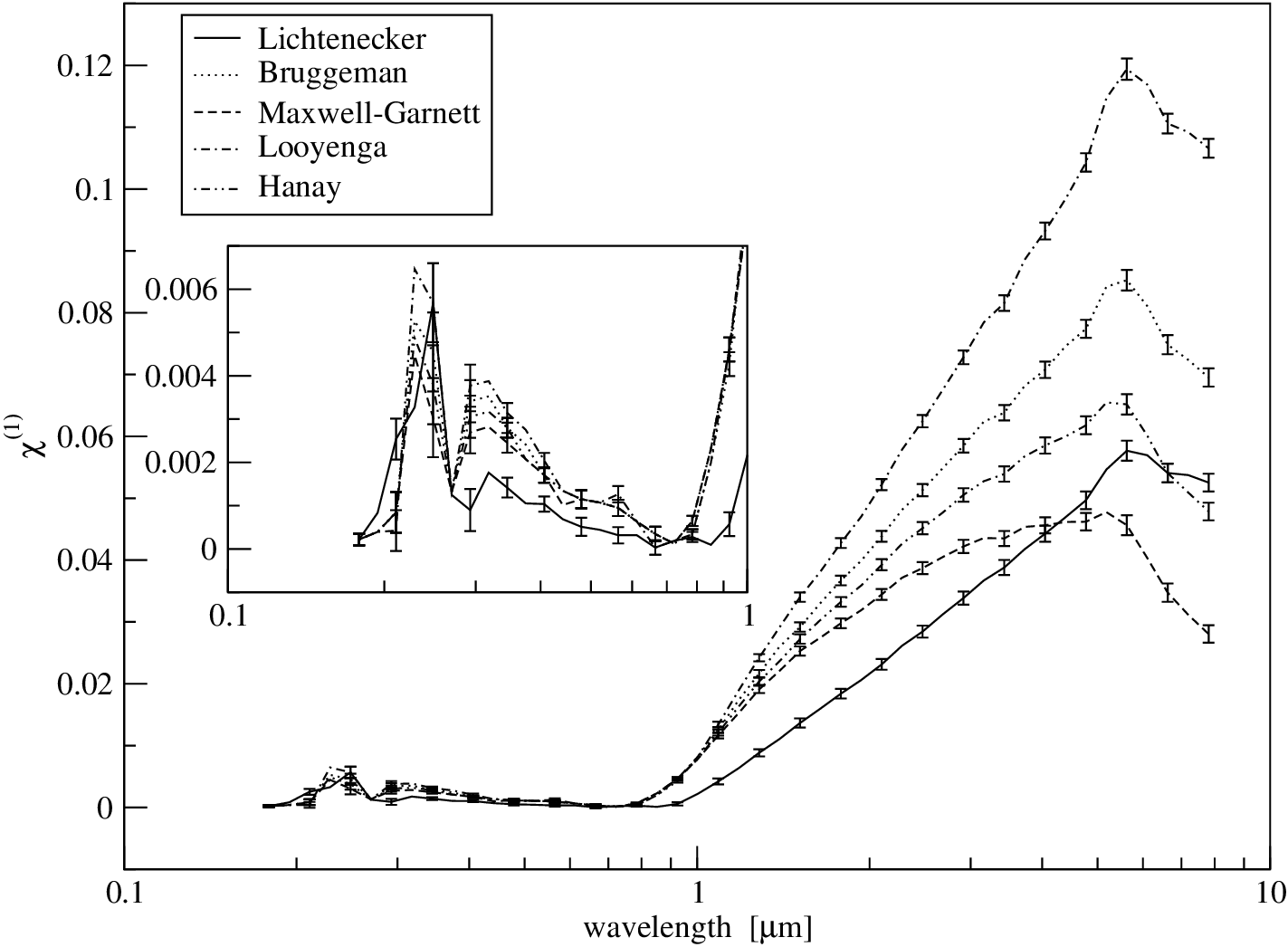} 
\caption{Example of dependance of $\chi^{(1)}$ on the wavelength with of error bars of standard deviation (see text)}
\label{errors}
\end{figure}

An example of the dependence of $\chi^{(1)}$ on the wavelength for the mixture of carbon (matrix) and graphite (inclusions) for 20\% of graphite inclusions with error bars $\Delta\chi^{(1)}$ is shown in  Figure~\ref{errors}.

In order to choose the best mixing rule in the whole considered range of wavelengths according to the Draine criteria we have calculated the goodness-of-fit parameter $\chi_{p,j}^{(2)}$ (further used as $\chi^{(2)}$)
\begin{equation}
\chi_{p,j}^{(2)}=\frac{1}{L}\sum_{l=1}^{L}\chi_{l,j,p}^{(1)},
\end{equation}
where L is the number of wavelengths for which the extinction coefficient has been calculated.

The standard deviation of $\chi_{p,j}^{(2)}$ for a given jth volume fraction is calculated as:
\begin{equation}
\Delta\chi_{p,j}^{(2)}=\frac{1}{L}\sum_{l=1}^{L}\Delta\chi_{l,j,p}^{(1)}
\end{equation}
and is shown as error bars in Figures~\ref{l_sil}-\ref{ice_vac} marked with letter f.

\begin{figure}
\includegraphics[width=84mm]{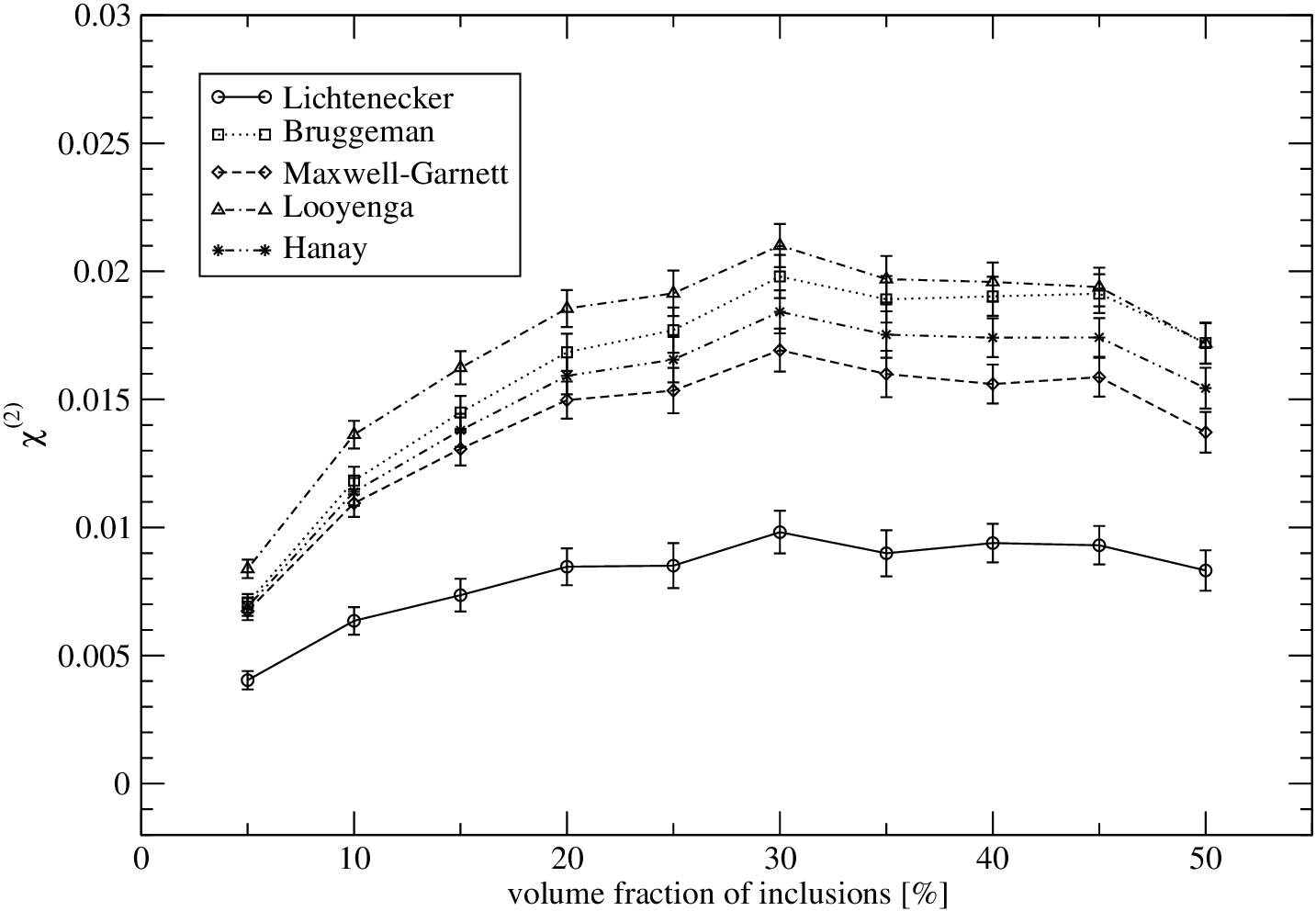} 
\caption{Goodness-of-fit parameter $\chi^{(1)}$ for scattering versus volume fractions of inclusions.}
\label{scatter}
\end{figure}

The dependence of $\chi^{(1)}$ on wavelength for the studied mixtures and volume fractions from 10\% to 50\% with 10\% step is shown in Figures~\ref{l_sil}-\ref{sil_vac} marked with letters a-e. The inspection of the figures allows to determine the best fit of mixing rule in different wavelength ranges with a given volume fraction.
In Figures~\ref{l_sil}-\ref{ice_vac} marked with letter f the best fit in the whole range of wavelengths depending on volume fraction of inclusions has been shown.

\begin{figure}
\includegraphics[width=84mm]{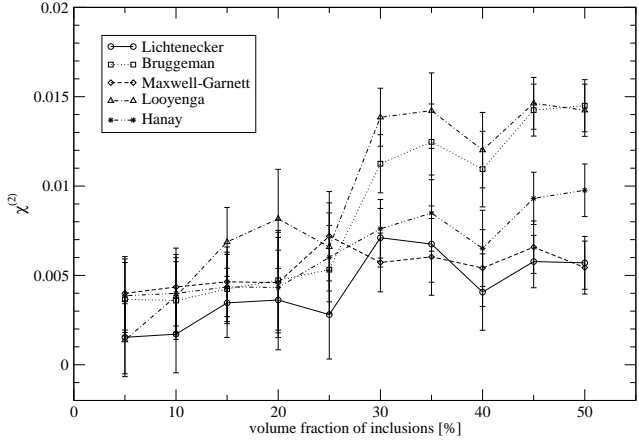} 
\caption{Goodness-of-fit parameter $\chi^{(2)}$ for asymmetry parameter g versus volume fractions of inclusions.}
\label{param-g}
\end{figure}

\begin{figure}
\includegraphics[width=84mm]{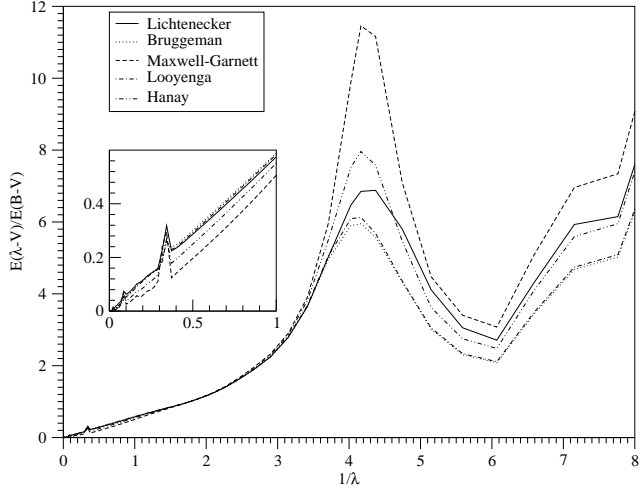} 
\caption{Example of influence of the used mixing rule on normalised extinction. Ice - matrix, graphite - inclusions (50\%)}
\label{extinc}
\end{figure}

The similar procedure as for extinction has been carried out for the scattering and the scattering asymmetry parameter $g\equiv<cos\theta>$. As an example we present the results of calculations of  $\chi_{p,j}^{(2)}$  with error bars for carbon (matrix) and graphite (inclusions) for scattering in Figure~\ref{scatter} and for asymmetry parameter in Figure~\ref{param-g}.
Compared to extinction the use of scattering does not improve the choice of the best mixing rule although in some, quite rare, cases it allows to choose better curves for some mixing rules. Furthermore,  in case of the parameter g the obtained results are much worse. In Figure~\ref{extinc} we have shown an example of influence of different mixing rules on the normalised extinction $E(\lambda-V)/E(B-V)$ calculated from Mie theory. The dependence on $1/\lambda$ is shown for grains with radius $0.02\mu m$ and consisting of ice (matrix) and 50\% graphite inclusions.

\section{Results and discussion}

For the symmetrical rules studied in this paper (Lichtenecker, Bruggeman, Looyenga) the values of the resulting permittivity of the mixture are the same for ${\rm f=50\%}$ both when the material A is an inclusion in the material B and vice versa. For the values of $\chi^{(1)}$ and $\chi^{(2)}$ there are slight differences for ${\rm f=50\%}$. It is caused by a different arrangement of inclusions when the material A is an inclusion in the material B than for an inverse situation (B - inclusion, A - matrix). Different arrangements of the same amount of inclusions give slightly different values of $Q_{l,j}^{rand}$.\\

\subsection{Mixture of "astronomical" silicate and ice}
We have considered the "astronomical" silicate inclusions in the icy matrix. For the volume fraction smaller than 10\% the best agreement of extinction coefficient of the mixture obtained from the mixing rule with the one of a mixture with "random" arrangement of inclusions in the whole wavelength range have been obtained for the Lichtenecker mixing rule - $\chi^{(1)}$ has the smallest value. The Lichtenecker rule $\chi^{(2)}$ is the smallest up to 17\% of volume fraction of inclusions. Above the 17\% the best rules are listed in Table~\ref{tab1} and it is seen that the differences in $\chi^{(2)}$ between Lichtenecker and Maxwell-Garnett rules are very small. These results are displayed in Fig.~\ref{l_sil}.

\begin{table}
\centering

\caption[]{Best mixing rules for the volume fraction of inclusions higher than 17\% for the mixture of "astronomical" silicate in ice matrix}
         \label{tab1}
          \begin{tabular}{@{}lc@{}}  
            \hline
                                 
	    Mixing rule   & $\chi^{(2)}_{max}$ \\
            \hline
            
            	    Maxwell$-$Garnett & 0.022\\
	    Lichtenecker (symmetrical) & 0.025 \\
	    Hanay &  0.033\\
	    Bruggeman (symmetrical) & 0.048 \\
	    Looyenga (symmetrical) & 0.058\\
            
            \hline
         \end{tabular}

\end{table}

In case of ice inclusions in "astronomical" silicate matrix for the volume fractions of inclusions up to 13\% the best is the Looyenga rule but not much worse is the Lichtenecker one - in both cases $\chi^{(2)}<0.01$. Above 13\% of volume fraction of inclusions the best rules are listed in Table~\ref{tab2} and shown in Fig.~\ref{sil_l}.

\begin{table}
\centering

\caption[]{Best mixing rules for the volume fraction of inclusions higher than 13\% for the mixture of ice in "astronomical" silicate matrix}
         \label{tab2}
          \begin{tabular}{@{}lc@{}}  
            \hline
                                 
	    Mixing rule   & $\chi^{(2)}_{max}$ \\
            \hline
            
            	    Lichtenecker (symmetrical) & 0.022\\
	    Looyenga (symmetrical) & 0.042 \\
	    Bruggeman (symmetrical) & 0.043 \\
	    Hanay &  0.059\\
	    Maxwell$-$Garnett & 0.071\\
            
            \hline
         \end{tabular}

\end{table}

\subsection{Mixture of ${\rm \alpha-SiC}$ and ice}
In case of ${\rm \alpha-SiC}$ inclusions in the ice matrix (Fig.~\ref{l_sic}) up to 20\% of volume fraction of inclusions the best rule is the Lichtenecker one. In the range from 20 to 40\% slightly better from the Lichtenecker is the Maxwell-Garnett mixing rule ($\Delta\chi^{(2)}<0.004$). Above 40\% again the best is Lichtenecker mixing rule.

For the inverse case (Fig.~\ref{sic_l}) the Looyenga rule is best for volume fraction of inclusions up to 12\% and in the range from 12\% to 50\% the Lichtenecker mixing rule gives the best results. Next in the order of goodness are Bruggeman, Hanay and Maxwell-Garnett. 
The best mixing rule for the mixtures of ${\rm \alpha-SiC}$  and ice is the Lichtenecker rule. For both cases of such mixtures the applicability of mixing rules has been examined for different wavelength ranges.

\subsection{Mixture of graphite and ice}
For the graphite inclusions in ice matrix up to 20\% of volume fraction of inclusions the best results gives the Maxwell-Garnett rule and above the 20\% the Hanay rule is best. Unfortunately both rules are asymmetrical  and therefore their applicability for high values of volume fraction of inclusions is limited.

In case of graphite matrix with ice inclusions in the whole volume fraction range the best results have been obtained for the Lichtenecker mixing rule. Figures~\ref{l_g}~and~\ref{g_l} show the results of calculations for the above cases.

\subsection{Mixture of carbon and ice}
Up to 35\% of volume fraction of carbon inclusions in the ice matrix the best is the Maxwell-Garnett rule and above that value the best is the Lichtenecker one. Next best rules are: Hanay, Bruggeman and Looyenga (Fig.~\ref{c_l}).

In the inverse case for the whole range of volume fractions the best results gives the Lichtenecker rule and next best rules are: Looyenga, Hanay and Maxwell-Garnett (Fig.~\ref{l_c}). For both cases of the mixture of ice and carbon considering the \citet{draine1988} criteria the applicability of the rules has been examined for different wavelength ranges.

\subsection{Mixture of Fe and ice}
Hanay (Bruggeman's asymmetric) formula corresponds best with experiment for large differences of complex permittivities between metallic inclusion and dielectric matrix \citep{merill1999}. This fact is confirmed for the inclusions of iron in ice (Fig.~\ref{l_fe}). \citet{merill1999} also stated on the basis of experiment that the Looyenga rule is only valid for low contrast between inclusion $\epsilon_{i}$ and matrix $\epsilon_{m}$ permittivity and thus is not appropriate in the metallic limit which is clearly seen in Fig.~\ref{l_fe}. The best mixing rules for this mixture are listed in Table~\ref{tab3}.
\begin{table}
\centering

\caption[]{Best mixing rules for Fe (inclusion) - ice (matrix)}
         \label{tab3}
          \begin{tabular}{@{}cc@{}}  
            \hline
                                 
            	      Volume fraction of inclusions [\%] & Mixing rule \\
        \hline
            	    0$-$10 & Maxwell$-$Garnett\\
	    10$-$33 & Hanay\\
	    33$-$50 & Lichtenecker\\
	              
            \hline
         \end{tabular}

\end{table}

The mixture of ice inclusions in iron matrix satisfies the \citet{draine1988} criteria for the 15\% to 50\% of ice content in iron matrix. In this range of volume fraction the best mixing rule is the formula by Lichtenecker and then down to the worst: Looyenga, Bruggeman, Hanay and Maxwell-Garnett (Fig.~\ref{fe_l}).

\subsection{Mixture of graphite and carbon}
In the whole range of volume fractions of graphite inclusions in carbon the best rule is the one by Lichtenecker and then in order of decreasing goodness: Maxwell-Garnett, Hanay and Looyenga (Fig.~\ref{c_g}). Below 10\% the quality factor for Maxwell-Garnett is slightly smaller than for the Lichtenecker rule.

For carbon inclusions in graphite matrix in the whole range of volume fractions the best rule is the one by Lichtenecker and then in order of decreasing goodness: Looyenga, Bruggeman, Hanay and Maxwell-Garnett (Fig.~\ref{g_c}).

\subsection{Mixture of SiC and carbon}
For the mixture of SiC inclusions in carbon matrix in the whole range of volume fractions of inclusions and wavelengths the best results have been obtained by using the Lichtenecker mixing rule. Next in decreasing order have been: Looyenga, Bruggeman, Hanay and Maxwell-Garnett (Fig.~\ref{sic_c}).

For the mixture of carbon inclusions in SiC matrix the Lichtenecker mixing rule gives the best results in the whole range of volume fractions of inclusions and wavelengths. Next in decreasing order have been: Maxwell-Garnett, Hanay, Bruggeman and Looyenga (Fig.~\ref{c_sic}).

\subsection{Porous structures}
We have considered the "astronomical silicate", carbon, iron and ice containing the vacuum pores which have been treated as inclusions of the same dimensions randomly distributed with refractive index $m=1.0+i1\cdot10^{-10}$. Calculations have been carried out in the same way as in previous cases with 50\% of inclusions (porosity).
Similar calculations for porous materials with refractive indices characteristic for "dirty ice", silicate and amorphous carbon in the visual wavelengths range have been carried out by \citet{voshchin2007}.

\subsubsection{Astronomical silicate with vacuum inclusions}
The Figure~\ref{sil_vac}~f shows that up to 31\% of volume fraction of inclusions the best results are obtained from the Looyenga rule and above that value from the Lichtenecker mixing rule. Next in the order of goodness are Bruggeman, Hanay and Maxwell-Garnett. 
For different volume fractions of inclusions in different wavelength ranges the best mixing rules are listed in Table~\ref{tab4} which is the summary of results shown in Figures~\ref{sil_vac}~a-e.

\begin{table}
\centering

\caption[]{The choice of best mixing rules depending on volume fraction of pores in silicate for different wavelengths}
         \label{tab4}
          \begin{tabular}{@{}crl@{}}  
            \hline
                                 
	       Volume fraction & wavelength  & Mixing rule\\
            of inclusions [\%] & range [${\rm \mu m}$] &  \\
        \hline
        10 & 0.2$-$10 & Looyenga \\
	         & 10$-$60   & Looyenga, Lichtenecker\\
	         & $>$60         & Bruggeman, Maxwell-Garnett\\
 \hline
	    20 & 0.2$-$10 & Looyenga \\
             & 10$-$50   & Lichtenecker\\
             & 50$-$100   & Looyenga\\
             & $>$100         & Bruggeman, Maxwell-Garnett\\
 \hline
        25 & 0.2$-$10 & Looyenga \\
	         & 10$-$60   & Lichtenecker\\
	         & $>$60         & Looyenga\\
 \hline
		30 & 0.2$-$10 & Looyenga \\
	         & 10$-$80   & Lichtenecker\\
	         & $>$80         & Looyenga\\
 \hline
		40 & $<$10 & Looyenga \\
	         & $>$10   & Lichtenecker\\
 \hline
		50 & $<$1.5 & Looyenga, Bruggeman \\
	         & 1.5$-$6   & Lichtenecker\\
	         & $>$10         & Lichtenecker\\

            \hline
         \end{tabular}
\end{table}

\subsubsection{SiC with vacuum inclusions}
From Figure~\ref{sic_vac}f it is seen that the dependance of ${\rm \chi^{(2)}}$ versus volume fraction of inclusions is similar to the same dependance for silicate with vacuum inclusions (Fig.~\ref{sil_vac}f). In the range up to 30\% the best results are obtained with Looyenga mixing rule. The other rules give almost the same quite good values of fit parameter ${\rm \chi^{(2)}}$ up to 20\% of volume fractions of inclusions. From 20\% to 30\% the best mixing rules are Looyenga, Lichtenecker,  Bruggeman, Hanay, Maxwell-Garnett. Above 30\% the best results are obtained from Lichtenecker rule and next from Looyenga, Bruggeman, Hanay and Maxwell-Garnett. The dependance of fitting parameter ${\rm \chi^{(1)}}$ on the wavelength for different mixing rules and volume fractions of inclusions are shown in Figures~\ref{sic_vac}~a-e.

\subsubsection{Graphite with vacuum inclusions}
The Fig.~\ref{g_vac}~f  shows that in the range from 5\% to 50\% of volume fraction of inclusions (pores) the best mixing rule is the Lichtenecker one. Next best are Looyenga, Bruggeman, Hanay and Maxwell-Garnett. For different volume fractions of inclusions in various wavelength ranges the best mixing rules have been listed in Table~\ref{tab6} which summarises the results shown in Figures~\ref{g_vac}~a-e.

\subsubsection{Carbon with vacuum inclusions}
From the Figure~\ref{c_vac}~f one can see that in the range from 5\% to 50\% of volume fraction of inclusion (pores) the best mixing rule is the Lichtenecker one. The next best are Looyenga, Bruggeman, Hanay and Maxwell-Garnett.
For different volume fractions of inclusions in different wavelength ranges the best mixing rules are listed in Table~\ref{tab5} which is the summary of figures Figures~\ref{c_vac}~a-e.

\begin{table}
\centering

\caption[]{The choice of best mixing rules depending on volume fraction of pores in carbon for different wavelengths}
         \label{tab5}
          \begin{tabular}{@{}crl@{}}  
            \hline
                                 
	       Volume fraction & wavelength  & Mixing rule\\
            of inclusions [\%] & range [${\rm \mu m}$] &  \\
        \hline
        10 & 0.2$-$0.5  & All rules \\
	         & 0.5$-$11   & Looyenga\\
	         & 11$-$150   & Lichtenecker\\
 \hline
	    20 & 0.2$-$6 & Lichtenecker \\
             & 6$-$10   & Looyenga\\
             & 10$-$40   & Hanay, Bruggeman, Maxwell-Garnett\\
             & $>$40         & Lichtenecker\\
 \hline
		30 & 0.2$-$10 & Lichtenecker \\
	         & 10$-$150   & Looyenga\\
 \hline
		40 & 0.2$-$0.75 & Looyenga \\
		     &0.75$-$30 & Lichtenecker\\
	         & 30$-$53   & Looyenga\\
			 & $>$53      & Bruggeman\\
 \hline
		50 & $<$0.6 & Looyenga \\
	         & 0.6$-$100   & Lichtenecker\\
	         & $>$100       & Maxwell-Garnett\\
       
            \hline
         \end{tabular}
\end{table}

\begin{table}
\centering

\caption[]{The choice of best mixing rules depending on volume fraction of pores in graphite for different wavelengths}
         \label{tab6}
          \begin{tabular}{@{}crl@{}}  
            \hline
                                 
	       Volume fraction & wavelength  & Mixing rule\\
            of inclusions [\%] & range [${\rm \mu m}$] &  \\
        \hline
        10 & 0.37$-$0.44  & Lichtenecker (best)\\
            &                      & Looyenga, Bruggeman, Maxwell-Garnett, Hanay \\
	         & 0.44$-$0.75   & Bruggeman, Maxwell-Garnett, Hanay (almost the same) \\
             &                    & Looyenga, Lichtenecker. For all rules $\chi<$0.01\\
	         & 0.75$-$1.85   & Lichtenecker (best)\\
             &                     & Looyenga, Bruggeman, Hanay, Maxwell-Garnett\\
             & 1.85$-$2.30   & Looyenga (best)\\
             &                     & Lichtenecker, Bruggeman, Hanay, Maxwell-Garnett\\
 \hline
	    20 & 0.37$-$0.41 & Lichtenecker (best) \\
             &                   & Looyenga, Bruggeman, Hanay, Maxwell-Garnett\\
             & 0.41$-$0.51   & Bruggeman, Maxwell-Garnett, Hanay (almost the same)\\
             &                     & Looyenga, Lichtenecker. For all rules $\chi<$0.01\\
             & 0.55$-$2.40   & Lichtenecker (best)\\
             &                     & Looyenga, Bruggeman, Hanay, Maxwell-Garnett\\
             & 2.40$-$2.70   & Looyenga (best), Lichtenecker\\
             &                     & Maxwell-Garnett, Hanay, Bruggeman (almost the same)\\
 \hline
		30 & 0.32$-$0.35 & Lichtenecker (best)\\
             &                   & Looyenga, Bruggeman, Hanay, Maxwell-Garnett\\
             & 0.35$-$0.50 & Bruggeman, Maxwell-Garnett, Hanay (almost the same)\\
             &                   & Looyenga, Lichtenecker. For all rules $\chi<$0.02\\
             & 0.52$-$2.50 & Lichtenecker (best)\\
             &                   & Looyenga, Bruggeman, Hanay, Maxwell-Garnett\\
             & 2.50$-$3.35 & Lichtenecker (best)\\
             &                   & Looyenga, Looyenga, Maxwell-Garnett, Bruggeman\\
 \hline
		40 & 0.16$-$0.45 & All rules give similar results, maximum $\chi(M-G)=0.054$\\
             & 0.45$-$2.5 & Lichtenecker (best)\\
             &                 & Looyenga, Bruggeman, Hanay, Maxwell-Garnett\\
             & 2.5$-$4     & Lichtenecker (best)\\
             &                   & Looyenga, Maxwell-Garnett, Hanay, Bruggeman\\
 \hline
		50 & 0.15$-$0.36 & All rules give similar results, maximum $\chi(M-G)=0.081$\\
            &                   & Lichtenecker (best)\\
            & 0.36$-$2.6 & Lichtenecker (best)\\
             &                 & Looyenga, Bruggeman, Hanay, Maxwell-Garnett\\
             & 2.6$-$3.7 & Lichtenecker (best)\\
             &                 & Looyenga, Bruggeman, Maxwell-Garnett, Hanay\\
              & 4.2$-$4.8 & Lichtenecker (best)\\
             &                 & Looyenga, Maxwell-Garnett, Bruggeman, Hanay\\
             & 4.8$-$5.2 & Lichtenecker (best)\\
             &                 & Maxwell-Garnett, Looyenga, Hanay, Bruggeman\\
       
            \hline
         \end{tabular}
\end{table}

\subsubsection{Iron with vacuum inclusions}
Due to  the \citet{draine1988} criteria the applicability of mixing rules is limited to narrow wavelength ranges especially for low volume fractions of inclusions. For example, the wavelength range for 10\% of volume fraction of inclusions is for ${\rm \lambda=0.48 \div 0.52 \mu m}$ and for 50\% ${\rm \lambda=0.16 \div 0.85 \mu m}$.

The Figure~\ref{fe_vac}~f  shows that in the range from 5\% to 50\% of volume fractions of vacuum inclusions the best results have been obtained using the Lichtenecker rule. Only in case of 50\% of volume fraction of inclusions below $ 0.4 \mu m$ better results gives the Bruggeman rule. Essentially all rules give equally good results.

\subsubsection{Ice with vacuum inclusions}
From  Fig.~\ref{ice_vac}~f  one can clearly see that in the whole range of volume fractions of inclusions the best results have been obtained from the Looyenga rule. Next in order of goodness are Bruggeman, Hanay, Lichtenecker and Maxwell-Garnett. In the range of very small values of k for ice (order of $10^{-8}$) only the Wiener criterion has been used because ice in this range acts as a very good dielectric.

\section{Conclusions}
Considering the presented results it is clearly seen that different mixing rules play important role depending on examined wavelengths. Nevertheless, from all chosen mixing rules for the considered materials in most cases (13 out of 20) the best results have been obtained using the Lichtenecker mixing rule. In 5 cases the Lichtenecker rule is best only for some volume fraction of inclusions. In case of graphite inclusions in ice the best mixing rule is the Maxwell-Garnett one for up to 20\% of volume fraction and Hanay for higher volume fractions. The Looyenga mixing rule gives best results for porous ice. In case of interstellar or circumstellar grains the processes leading to their nucleation and growth are complicated and depending on not always known, changing in time and space physical and chemical conditions. It causes difficulties in determining the constituents of grains, their topology, shape and parameters of their size distribution. The choice of the best mixing rule decreases the number of free parameters when choosing the computed extinction curves to compare with the interstellar or circumstellar extinction. Components of the mixtures for which the best results are obtained from  Lichtenecker rule may form the multicomponent grains, as for example proposed by \citet{mathis1989}, with better derived effective dielectric function. Therefore, it is justified to apply the mixing rules chosen based on the numerical experiments.

\section*{Acknowledgements}
We thank Dr. Michael J. Wolff for his useful comments and constructive suggestions which greatly improved this paper. O.M. acknowledges a partial support by Polish Grant No. N  N203 2738 33.

\begin{figure}
\includegraphics[width=164mm]{fig_g_l_all4.eps} 
\caption{Graphite - matrix, ice - inclusions (same as in Fig.~\ref{l_sil})}
\label{g_l}
\end{figure}

\begin{figure}
\includegraphics[width=164mm]{fig_l_fe_all4.eps} 
\caption{Ice - matrix, Fe - inclusions (same as in Fig.~\ref{l_sil})}
\label{l_fe}
\end{figure}

\begin{figure}
\includegraphics[width=164mm]{fig_g_c_all4.eps} 
\caption{Graphite - matrix, carbon - inclusions (same as in Fig.~\ref{l_sil})}
\label{g_c}
\end{figure}

\begin{figure}
\includegraphics[width=164mm]{fig_c_vac_all4.eps} 
\caption{Carbon - matrix, vacuum - inclusions (same as in Fig.~\ref{l_sil})}
\label{c_vac}
\end{figure}
\clearpage

\label{lastpage}

\begin{thebibliography}{}

\bibitem[\protect\citeauthoryear{Aspens}{1982}]{aspens1982}Aspens D. E., Am. J. Phys., 1982, 50, 8
\bibitem[\protect\citeauthoryear{Beek}{1967}]{beek1967}Beek van L. K. H., 1967, "Dielectric Behaviour of Heterogeneous Systems" in Progress in Dielectrics Vol. 7, London Heywood Books, ed. J. B. Birks
\bibitem[\protect\citeauthoryear{Bergman}{1980}]{bergman1980}Bergman D. J., Phys. Rev. Lett., 1980, 44, 1285
\bibitem[\protect\citeauthoryear{Bohren \& Huffman}{1983}]{bohren1983}Bohren C. F., Huffman D. R., 1983, Absorption and Scattering of Light by Small Particles, John Wiley \& Sons Inc., Toronto
\bibitem[\protect\citeauthoryear{Bohren \& Wickramasinghe}{1977}]{bohren1977}Bohren C. F., Wickramasinghe N. C., 1977, Astrophys. Space Sci., 50, 461
\bibitem[\protect\citeauthoryear{Bruggeman}{1935}]{bruggeman1935}Bruggeman, D. A. G., 1935, Ann. Phys. Lpz., 24, 636
\bibitem[\protect\citeauthoryear{Chylek \& Srivastava}{1983}]{chylek1983}Chylek, P., Srivastava V.,1983, Phys. Rev. B, 27, 5098
\bibitem[\protect\citeauthoryear{Chylek et al.}{2000}]{chylek2000}Chylek, P., Videen, G., Geldart, D.J.W., Dobbie J.S., Tso, H.C.W., 2000, in Light Scattering by Nonspherical Particles: Theory, Measurements, and Applications, Academic Press, p. 273
\bibitem[\protect\citeauthoryear{Draine}{1985}]{draine1985}Draine B. T., 1985, ApJ Suppl. Ser. 57, 587
\bibitem[\protect\citeauthoryear{Draine}{1988}]{draine1988}Draine B. T., 1988, ApJ 333,848
\bibitem[\protect\citeauthoryear{Draine \& Flatau}{1994}]{draine1994}Draine B. T., Flatau P. J., 1994, J. Opt. Am. A., 11, 1491
\bibitem[\protect\citeauthoryear{Draine \& Flatau}{2000a}]{draine2000}Draine B. T., Flatau P. J., 2000a,  http://www.astro.princeton.edu/$\sim$draine
\bibitem[\protect\citeauthoryear{Draine \& Flatau}{2000b}]{draine2000b}Draine B. T., Flatau P. J., 2000b,  http://xxx.lanl.gov/abs/astro-ph/00081151v3 (User Guide for DDA Code DDSCAT (Version 5a10)
\bibitem[\protect\citeauthoryear{Draine \& Goodman}{1993}]{draine1993}Draine B. T., Goodman J, 1993, ApJ, 405, 685
\bibitem[\protect\citeauthoryear{Iati et al.}{2004}]{iati2004}Iati M. A., Giusto A., Saija R., Borghese F., Denti P., Cechi-Pestellini C. and Aielo S., 2004, ApJ, 615, 286
\bibitem[\protect\citeauthoryear{Jones}{1988}]{jones1988}Jones, A.P., 1988, MNRAS, 234, 209
\bibitem[\protect\citeauthoryear{Lichtenecker}{1926}]{lichtenecker1926}Lichtenecker, K., 1926, Physikalische Zeitschrift, 27, 115
\bibitem[\protect\citeauthoryear{Looyenga}{1965}]{looyenga1965}Looyenga, H., 1965, Physica, 31, 401
\bibitem[\protect\citeauthoryear{Landau \& Lifszic}{1960}]{landau1960}Landau L. D., Lifszic E. M., 1960, Elektrodynamika o\'{s}rodk\'{o}w ci\c{a}g\l ych, PWN, Warszawa
\bibitem[\protect\citeauthoryear{Landauer}{1952}]{landauer1952}Landauer R., 1952, J. Appl. Phys., 23, 779
\bibitem[\protect\citeauthoryear{Lynch \& Hunter}{1991}]{lynch1991} Lynch D.W., Hunter W.R., 1991, in Handbook of Optical Constants of Solids II, Palik E.D., Academic Press
\bibitem[\protect\citeauthoryear{Merill et al.}{1999}]{merill1999}Merill, W. M., Diaz, R. E., Lore, M. M., Squires, M. C., Alexopulos, N. G., 1999, IEEE Transactions on Antennas and Propagation, vol. 47, no 1, 47, 142
\bibitem[\protect\citeauthoryear{Milton}{1980}]{milton1980}Milton G.W., Appl., Phys. Lett., 1980, 37, 3
\bibitem[\protect\citeauthoryear{Maron}{1989}]{maron1989}Maron, N., 1989, Astrophys., Space Sci., 161, 201
\bibitem[\protect\citeauthoryear{Maron \& Maron}{2005}]{maron2005}Maron, N., Maron, O., 2005, MNRAS, 357, 873
\bibitem[\protect\citeauthoryear{Mathis \& Whiffen}{1989}]{mathis1989}Mathis, J.S., Whiffen, G., 1989, ApJ, 341, 808
\bibitem[\protect\citeauthoryear{Maxwell-Garnett}{1904}]{mg1904}Maxwell-Garnett J.C., 1904, Philosophical Transactions of the Royal Society London Series A, Vol. 203, 385
\bibitem[\protect\citeauthoryear{Meredith \& Tobias}{1960}]{meredith1960}Meredith R. E., Tobias C. W., 1960, J. Appl. Phys., 31, 1270
\bibitem[\protect\citeauthoryear{Perrin \& Lamy}{1990}]{perrin1990}Perrin J.-M., Lamy P.L., 1990, ApJ, 364, 146
\bibitem[\protect\citeauthoryear{Sihvola}{1973}]{sihvola1999}Sihvola A., 1999, Electromagnetic mixing formulas and applications, IEE, London, p. 168
\bibitem[\protect\citeauthoryear{Purcell \& Pennypacker}{1973}]{purcell1973}Purcell E. M., Pennypacker C. R., 1973, ApJ, 186, 705
\bibitem[\protect\citeauthoryear{Vaidya et al.}{2001}]{vaidya2001}Vaidya, D.B., Gupta, R., Dobbie, J.S., Chylek, P., 2001, A\&A, 375, 584
\bibitem[\protect\citeauthoryear{Voshchinnikov \& Mathis}{1999}]{voshchin1999}Voshchinnikov, N.V., Mathis, J.S., 1999, ApJ, 526, 257
\bibitem[\protect\citeauthoryear{Voshchinnikov et al.}{2007}]{voshchin2007}Voshchinnikov, N.V., Videen G. and Henning T., 2007, Appl. Opt., 46, 4065
\bibitem[\protect\citeauthoryear{Warren}{1984}]{warren1984}Warren S.G., 1984, Appl. Opt., vol 23, no 8, 1206
\bibitem[\protect\citeauthoryear{Wolff et al.}{1994}]{wolff1994}Wolff M.J., Clayton G.C., Martin P.G. and Schulte-Ladbeck R.E., 1994, ApJ, 423, 412
\bibitem[\protect\citeauthoryear{Wolff et al.}{1998}]{wolff1998}Wolff M.J., Clayton G.C. and Gibson S.J., 1998, ApJ, 503, 815
\bibitem[\protect\citeauthoryear{Zakri et al.}{1998}]{zakri}Zakri T., Laurent J.-P., Vauclin M., 1998, Appl. Phys., 31, 1589
\bibitem[\protect\citeauthoryear{Zubko et al.}{1996}]{zubko1996}Zubko V.G., Mennella V., Colangeli L., Bussoletti E., 1996, MNRAS, 282, 1321
\end{thebibliography}
\end{document}